\documentclass[12pt]{article}
\usepackage{subfiles}

\usepackage[utf8]{inputenc}
\usepackage[T1]{fontenc}
\usepackage{bm}
\usepackage{times}  

\usepackage[a4paper, left=0.8in, right=0.6in, top=0.7in, bottom=0.7in]{geometry}
\usepackage{setspace}
\usepackage[left]{lineno} 

\usepackage{graphicx}
\usepackage{float}
\usepackage{wrapfig}
\usepackage[export]{adjustbox}
\usepackage{caption}
\usepackage{subcaption}

\usepackage{amsmath, amsfonts}
\usepackage{mathtools}
\usepackage{bbm}
\usepackage{dsfont}

\usepackage{array}
\newcolumntype{L}[1]{>{\raggedright\arraybackslash}p{#1}}

\usepackage{enumitem}
\setlist{
    topsep=0pt,
    partopsep=0pt,
    itemsep=0pt,
    parsep=0pt
}

\usepackage{algorithm}
\usepackage{algpseudocode}

\usepackage{booktabs}
\usepackage{makecell}
\usepackage{longtable}
\usepackage{array, multirow, boldline}
\usepackage{tabularx}
\usepackage[table]{xcolor}
\usepackage{multirow}

\usepackage{hyperref}
\usepackage{xurl}
\usepackage{url}
\usepackage{footnote}
\makesavenoteenv{tabular}
\makesavenoteenv{table}

\usepackage[backend=bibtex,style=numeric-comp, sorting=none, url=false, eprint=false, isbn=false, minnames = 3, maxnames=3,giveninits=true]{biblatex}
\AtEveryBibitem{%
  \clearfield{note}%
}

\usepackage{hyperref}

\usepackage{microtype}
\usepackage{blindtext}
\usepackage{comment}
\usepackage{listings}

\usepackage{tikz}
\usetikzlibrary{bayesnet}

\tikzstyle{arrow} = [thick, ->, >=stealth]
\tikzstyle{flowbox} = [
    rectangle,
    draw,
    thick,
    minimum width=3.8cm,
    minimum height=1cm,
    text centered,
    text width=3.7cm,
    align=center,
    fill=white
]
\tikzstyle{input} = [flowbox, draw=blue!60, fill=blue!10]
\tikzstyle{output} = [flowbox, draw=blue!60, fill=blue!10]
\tikzstyle{greenblock} = [flowbox, draw=green!60!black, fill=green!10]
\tikzstyle{redblock} = [flowbox, draw=red!70!black, fill=red!10]
\tikzstyle{yellowblock} = [flowbox, draw=orange!90!black, fill=yellow!15]

\usepackage{hyperref}
\usepackage{cleveref}

\title{
Adaptive Bayesian computation for efficient biobank-scale genomic inference
}

\author{
Yiran Li$^{1,}$\thanks{Corresponding authors: yiran.li@mrc-bsu.cam.ac.uk; helene.ruffieux@mrc-bsu.cam.ac.uk},
John Whittaker$^{1}$,
Sylvia Richardson$^{1}$,
Hélène Ruffieux$^{1,*}$\\[1em]
$^{1}$MRC Biostatistics Unit, University of Cambridge
}

\begin{document}

\maketitle
\begin{abstract}
\noindent

\textbf{Background:} Modern biobanks, with unprecedented sample sizes and phenotypic diversity, have become foundational resources for genomic studies, enabling powerful cross-phenotype and population-scale analyses. As studies grow in complexity, Bayesian hierarchical models offer a principled framework for jointly modeling multiple biological dimensions or ``units'', such as cells, traits and experimental conditions, increasing statistical power through information sharing. However, adoption of such models in biobank-scale studies remains limited due to computational inefficiencies, particularly in posterior inference over high-dimensional parameter spaces. Deterministic approximations such as variational inference provide scalable alternatives to Markov Chain Monte Carlo, yet current implementations do not fully exploit the structure of genome-wide multi-unit modeling, especially when biological effects are sparse, i.e., concentrated in a small number of units.

\textbf{Results:} We present AF-CAVI, an adaptive focus (AF) scheme within a block coordinate ascent variational inference (CAVI) framework that selectively updates subsets of parameters at each iteration, corresponding to units deemed relevant based on current estimates. We also develop a genome-wide multi-response protein quantitative trait loci (pQTL) mapping pipeline based on AF-CAVI–accelerated joint Bayesian modeling, and demonstrate its performance on simulated data and selected regions of UK Biobank chromosome 1. AF-CAVI achieves a $\approx 50\%$ runtime reduction compared to state-of-the-art CAVI with maintained statistical performance and improved power over univariate mapping. The AF-CAVI algorithm for pQTL mapping based on AF-CAVI-accelerated BHM is provided in a standalone R package called \texttt{AFatlasQTL}.

\textbf{Conclusions:} AF-CAVI makes joint Bayesian modeling practical for biobank-scale multi-trait inference by controlling computational cost while retaining statistical performance. More broadly, it provides a flexible computational framework for high-dimensional multi-unit models with sparse signals, supporting scalable genomic discovery across traits and conditions. 

\end{abstract}


\section{Background}


Modern biobanks, such as the UK Biobank \cite{ollier_uk_2005} and others participating in Global Biobank Meta Analyses \cite{zhou_global_2022}, offer exceptionally large sample sizes and rich phenotypic and genomic data, enabling analyses across a large spectrum of traits and populations. As such, biobanks have become foundational resources in modern genomics, facilitating a wide range of applications including genome-wide association studies (GWAS) \cite{bycroftUKBiobankResource2018}, quantitative trait locus (QTL) mapping \cite{sun_plasma_2023}, protein–protein interaction analysis \cite{suhre_genetic_2024} and, more recently, single-cell resolved transcriptional profiling \cite{sukhatme_integration_2024}. Such studies usually aim to detect associations between different omics layers, such as DNA variants, transcriptomics, proteomics, epigenomics, as well as their link with a broad array of phenotypic outcomes.  

Bayesian hierarchical models (BHMs) provide a powerful framework for the joint analysis of biobank data. They can capture complex relationships by jointly analyzing multiple \emph{units}, such as traits, tissues, cells or experimental conditions, and have been applied in a variety of contexts. For example, \textcite{luo_bayesian_2011} jointly modeled protein-protein interactions across different experimental levels of single-cell interventions within a Bayesian framework. In the context of QTL mapping, instead of conducting association tests for each genetic variant and trait separately \emph{(univariate association screening)}, Bayesian models are applied to jointly detect QTL effects for multiple traits \cite{banerjee_bayesian_2008} or to combine data from different tissues and cell types \cite{flutre_statistical_2013,cheng_hbi_2024}. One fundamental aspect of hierarchical models, such as \textit{hierarchically linked regressions}, is the use of \emph{local} parameters specific to each unit (or regression) and \emph{global} parameters shared across units, which are involved in the prior specification of the local parameters. Global parameters help to learn the overall characteristics of the problem such as sparsity levels, typical effect sizes and noise levels, while local parameters define the specific characteristics of the unit in question. Such a structure can largely increase statistical power in particular when some weak effects are shared. 



However, Bayesian hierarchical inference for genomics data usually involves obtaining complex, high-dimensional posterior distributions, which requires large computational resources. It is therefore rarely applied to biobank-scale data analysis, despite its relevance for uncovering biologically meaningful patterns with enhanced statistical power. Recent developments in scalable Bayesian inference have brought deterministic approximations to the forefront, providing a more efficient solution than traditional sampling-based methods like Markov Chain Monte Carlo (MCMC). In particular, variational inference (VI) algorithms approximate the posterior using simpler families of distributions, such as fully-factorized distributions in the case of mean-field VI \cite[see, e.g.,][]{blei_variational_2017}. Under global-local Bayesian model hierarchies, mean-field VI usually approximates the posterior into a product of \emph{factors} corresponding to the local and global structure. When the dimensionality is high (i.e., when the number of units is large, which is typical in genomics), a large number of local parameters need updating in turn, while holding the others fixed. When the sample size is large, some updates will also be computationally expensive. Therefore, although VI-based algorithms have made Bayesian inference computationally feasible for relatively large problems, they remain insufficiently nimble for processing genome-wide data in today's biobanks.

Current VI implementations of hierarchical models also often overlook an important feature defining most omics problems: while the total number of modeled units may be large, meaningful biological signals are typically sparse, namely, concentrated in only a small subset of \emph{``active''} local units. Exploiting this sparsity, adaptively focusing update efforts on these active units (in a way we are going to make precise below) could improve the algorithm's scalability without sacrificing accuracy. Following this observation, we propose an \emph{adaptive focus} (AF) strategy for the coordinate ascent approach of VI (CAVI), designed to enhance the scalability of BHMs for sparse, high-dimensional problems. At each iteration, the algorithm evaluates the activity of each unit based on a criterion (using the current posterior estimates), and accordingly defines an \emph{activity score} for each unit. This score is dynamically adjusted throughout the algorithm, leading to an \emph{adaptive selection process} that updates only a subset of parameters corresponding to active units. In this paper, we formulate the concept of AF-CAVI under a general global-local framework, and provide an example of AF-CAVI implementation for protein quantitative trait locus (pQTL) mapping and particularly for the detection of hotspots (loci associated with many traits) using the UK Biobank Pharma Proteomics Project (UKB-PPP) data \cite{sun_genetic_2022}. The AF-CAVI algorithm for pQTL mapping based on AF-CAVI-accelarated BHM is provided in a standalone R package called \texttt{AFatlasQTL}.



\section{Methods}


To present AF-CAVI as a framework for accelerating inference in BHM, we begin in  Section~\ref{sec:motivation_gl} by formalizing the key concepts of ``units'', and ``global'' and ``local'' components, using hierarchical regression as a motivating example. We then abstract these ideas to a broader class of BHMs, highlighting the structural properties that enable adaptive computation. Following that, we introduce the CAVI algorithm for BHM Section~\ref{sec:cavi}, establishing the optimization framework and its properties. In Section~\ref{sec:partial}, we then present our adaptive focus scheme to accelerate CAVI computation. Finally, in Section~\ref{sec:related work}, we provide a comparison of AF-CAVI with other acceleration strategies in Bayesian computation, such as stochastic VI and adaptive MCMC, and further interpret the method from a convex optimization perspective.

\subsection{Global-local Bayesian hierarchical models} \label{sec:motivation_gl}




Bayesian hierarchical regressions are widely applied for association studies in genomics. Consider a single-outcome problem with a response vector $\bm{y} \in \mathbb{R}^n$ and a set of $p$ candidate predictors collected in a design matrix $\bm{X} \in \mathbb{R}^{n \times p}$, for $n$ samples. Many epidemiological and biological problems reduce to estimating the effects of the predictors on the outcome, represented by regression coefficients $\beta_1,\ldots,\beta_p$, and selecting a parsimonious subset that contributes to explaining the variance in $\bm y$, a task known as \emph{variable selection}. Hierarchical regression provides a natural mechanism for regularization by modeling the regression coefficients as random variables governed by higher-level parameters, thereby enabling information sharing and adaptive shrinkage \cite{greenland_hierarchical_1994}. 

This idea extends naturally to settings with multiple responses \cite{richardson_hierarchical_2015}. Specifically, given $q$ responses observed for $n$ samples,  $\bm{y}_1,\ldots,\bm{y}_q \in \mathbb{R}^n$, each response $\bm{y}_t$ can be modeled as
\[
\bm{y}_t \mid \boldsymbol{\beta}_t, \tau_t 
\sim \mathcal{N}_n\!\left(\bm{X}\boldsymbol{\beta}_t,\; \tau_t^{-1}\mathbf{I}_n\right),
\quad t = 1,\ldots,q,
\]
where $\boldsymbol{\beta}_t = (\beta_{1t},\ldots,\beta_{pt})$ denotes the regression coefficient vector for response $t$, and $\tau_t$ is a response-specific residual precision parameter. In the Bayesian setting, a prior is placed on the collection of regression coefficient vectors $\{\boldsymbol{\beta}_t\}_{t=1}^q$, jointly modeling them through a multivariate distribution with shared parameters. This can induce shrinkage towards a common mean while maintaining sufficient flexibility for deviations from it given evidence in the data \cite{richardson_hierarchical_2015}. For example, when $p$ is large and the signal is assumed to be sparse, spike-and-slab prior \cite{ishwaran_spike_2005} assign non-negligible posterior probability only to a small subset of predictors, which enables simultaneous coefficient estimation and variable selection \cite{ohara_review_2009, griffin_hierarchical_2017}. Specifically, it consists of a mixture distribution with a point mass at zero, representing null effects (the spike), and a diffuse continuous distribution (can be a Gaussian with large variance, for example), representing non-null effects (the slab). In the multi-response setting, for each response $t$ we introduce a latent inclusion indicator vector $\boldsymbol{\gamma}_t = (\gamma_{1t},\ldots,\gamma_{pt})$, where $\gamma_{st} \in \{0,1\}$ indicates whether predictor $s$ is ``active'' for response $t$, i.e., has a nonzero coefficient:
\[
\beta_{st} \mid \gamma_{st}, \tau_t, \sigma \sim \gamma_{st} \mathcal{N}(0, \sigma^2 \tau_t^{-1}) + (1 - \gamma_{st}) \delta_0, \quad s = 1, \ldots, p, 
\] 
where $\delta_0$ is the Dirac delta distribution and $\sigma$ controls the overall scale of non-zero effects.  Figure~\ref{fig:model} provides a graphical representation of the model.

\begin{figure}[ht] 
\centering
\begin{tikzpicture} 

  \node[latent] (t_t) {$\tau_t$};
  \node[latent, below left=of t_t, xshift=-0.1cm, yshift=-0.2cm] (b_st) {$\bm{\beta}_{t}$};
  \node[obs, below right=of t_t, xshift=0.1cm, yshift=-0.2cm] (y_t) {$\bm{y}_t$};
  \node[obs, right=of y_t, xshift=0.8cm] (X) {$\bm{X}$};
  \node[latent, below=of b_st] (s) {$\sigma$};
  \node[latent, left=of b_st, xshift=-0.15cm] (g_st) {$\bm{\gamma}_{t}$};

  \plate[inner sep=0.35cm] {plate1} {(t_t)(g_st)(b_st)(y_t)} {$t = 1, \dots, q$};

  \edge {g_st} {b_st};
  \edge {b_st} {y_t};
  \edge {t_t} {y_t};
  \edge {t_t} {b_st};
  \edge {X} {y_t};
  \edge {s} {b_st}

\end{tikzpicture}
\caption{\label{fig:model} \textbf{Hierarchical multi-response regressions with spike-and-slab prior for the regression coefficients.} The shaded nodes are observed, the others are inferred.  Each regression $t$ relates candidate predictors $\bm X\in \mathbb{R}^{n \times p}$ to a response $\bm y_t\in \mathbb{R}^{n}$.  The regression coefficient vector $\boldsymbol{\beta}_t$, latent binary indicator vector $\boldsymbol{\gamma}_t$ and residual  precision $\tau_t$ 
are local to regression $t$, whereas $\sigma$ is a global variance parameter shared across regressions. }
\end{figure}


Besides the spike-and-slab prior, a broad class of discrete and continuous shrinkage priors can be employed for the regularization of regression coefficients \cite{park_bayesian_2008}. Further levels of hierarchy can also be introduced to accommodate latent variables and conditional dependencies across regressions \cite{de_jong_hierarchical_1999}. Hierarchical formulations also extend beyond regression models to a range of probabilistic settings in genomics and systems biology \cite{bae_gene_2004, gompert_hierarchical_2011}. The defining feature is the explicit organization of parameters into levels, each comprising a collection of \emph{units}  (e.g., predictors, responses, or experimental contexts), which enables principled sharing of statistical strength across related estimation problems while preserving unit-specific effects and uncertainty quantification. Parameters referring to individual units, hereafter termed \emph{local} parameters, determine unit-specific characteristics, whereas \emph{global} parameters govern their joint prior distribution, thereby coupling these units and controlling the extent of shrinkage and dependence across units. In the multi-response regression setting considered here (Figure~\ref{fig:model}), each response $\bm y_t$, for $t \in \{1, \ldots, q\}$, constitutes a unit; parameters $\boldsymbol{\beta}_t$, $\boldsymbol{\gamma}_t$ and $\tau_t$ capture response-specific effect sizes and noise levels and are therefore local, whereas $\sigma$ acts as a global parameter controlling the overall variability of regression effects across different responses.

More generally, many hierarchical models admit a representation in which observations and parameters are indexed by units and coupled through shared global parameters. Suppose the observations can be grouped into $q$ units, $\bm{y}_1,\ldots,\bm{y}_q$, and write $\bm{y}=(\bm{y}_1,\ldots,\bm{y}_q)$. Let $\bm{\nu}$ denote the full set of model parameters and introduce the notation $\bm{\nu}=(\bm{g},\bm{\ell})$, where $\bm{g}$ collects the global parameters and $\bm{\ell}=(\bm{\ell}_1,\ldots,\bm{\ell}_q)$ the local parameters with $\ell_t$ referring to unit $t$. A common hierarchical construction assumes that, conditional on the global parameters, (i) local parameters are mutually independent across units, and (ii) each observation depends only on its corresponding local parameters (and possibly $\bm g$). This conditional independence structure arises naturally, for example, when $\bm g$ represents shared hyperparameters governing unit-specific regression coefficients, variances or latent effects. 
Under these assumptions, the joint distribution factorizes as
\begin{equation} \label{formula:global-local joint prob}
p(\bm{y}, \bm{\nu}) =
p(\bm{g})
\prod_{t=1}^q
p(\bm{\ell}_t \mid \bm{g})
p(\bm{y}_t \mid \bm{\ell}_t, \bm{g}),
\end{equation}
which implies, by Bayes' theorem and conditional independence, the following ``global-local'' posterior distribution 
\begin{equation} \label{formula:global-local posterior}
p(\bm{\nu}\mid \bm{y}) = p(\bm{g}\mid \bm{y})\prod_{t=1}^q p(\bm{\ell}_t\mid \bm{g},\bm{y}_t).
\end{equation}
This factorization separates inference on the shared global parameters from a collection of conditionally independent unit-specific inference problems. Such structure is characteristic of hierarchical models with exchangeable units and provides a natural structure for scalable Bayesian computation and optimization strategies, as we detail in the next section.





\subsection{CAVI under the global-local decomposition} \label{sec:cavi}
Variational inference approximates the posterior $p(\boldsymbol{\nu \mid y})$ by a tractable \emph{variational distribution} $q(\boldsymbol{\nu})$, chosen to  minimize the reverse Kullback–Leibler (KL) divergence:
\begin{flalign} \nonumber
\text{KL} \left[q (\bm{\nu}) \,\Vert\, p(\bm{\nu} \mid \bm{y}) \right] 
&= \mathbb{E}_{q(\bm{\nu})}[\log q(\bm{\nu})]
  - \mathbb{E}_{q(\bm{\nu})}[\log p(\bm{\nu} \mid \bm{y})] \\ \nonumber
&= -\mathbb{E}_{q(\bm{\nu})} \left[ \log \frac{p(\bm{y}, \bm{\nu})}{q(\bm{\nu})} \right]
  + \log p(\bm{y}),
\end{flalign}
where $\mathbb{E}_{q(\bm\nu)}$ denotes expectation with respect to  $q(\bm \nu)$. Since the marginal log-evidence $\log p(\bm{y})$ does not depend on $q$, minimizing the KL divergence with respect to $q$ is equivalent to maximizing the \emph{evidence lower bound} (ELBO), defined as
\begin{equation} \label{formula:elbo}
\begin{aligned}
    \mathcal{L}(q(\bm{\nu})) &= \mathbb{E}_{q(\bm{\nu})}\left[\log \frac{p(\bm{\nu} , \bm{y})}{q(\bm{\nu})}\right] \\
    &= \mathbb{E}_{q(\bm{\nu})}[\log p(\bm{y}, \bm{\nu})] - \mathbb{E}_{q(\bm{\nu})}[\log q(\bm{\nu})].
\end{aligned}
\end{equation} 
Maximizing $\mathcal{L}(q(\bm{\nu}))$ yields the closest approximation to the posterior within the chosen variational family while avoiding direct evaluation of the intractable log-evidence. The first term in~\eqref{formula:elbo} corresponds to the expected log joint distribution, while the second term is the entropy of the variational distribution. 


A useful family of variational distributions employed to approximate $p(\bm{\nu \mid \bm{y}})$ in BHMs is the mean-field variational family, where $q(\boldsymbol{\nu})$ is decomposed into \emph{factors} that are assumed independent of each other and each governed by a distinct variational parameter vector (also referred to as a block of parameters) $\bm{\nu}_j$:
\[
q(\boldsymbol{\nu}) = \prod_j q(\bm{\nu}_j) .
\]

Reflecting the conditional independence structure in \eqref{formula:global-local posterior}, the following mean-field factorization of the variational distribution naturally arises:
\begin{equation}\label{formula:mean-field decomp}
    q(\boldsymbol{\nu}) = q(\bm{g})  \prod_{t=1}^q q(\bm{\ell}_t) , 
\end{equation} 
where each \emph{local factor} $q(\bm{\ell}_t)$ corresponds to the variational approximation for the posterior $p(\bm{\ell}_t \mid\bm{y})$ of model unit~$t$, while the \emph{global factor} $q(\bm{g})$ is an approximation for the posterior $p(\bm{g} \mid \bm{y})$. 


Under conditional conjugacy, coordinate ascent variational inference (CAVI) updates each factor by maximizing the ELBO with respect to one block at a time in the form of:
\[
\log q(\boldsymbol{\nu}_j) = \mathbb{E}_{q(\bm{\nu}_{-j})}[\log p(\bm{y}, \boldsymbol{\nu})] +\mathrm{const},
\]
where $\mathbb{E}_{q(\bm\nu_{-j})}$ denotes expectation with respect to all variational factors except $q(\bm \nu_j)$ and $\mathrm{const}$ is a constant with respect to $\bm \nu_j$. With decomposition \eqref{formula:mean-field decomp} of $q(\bm{\nu)}$ and decomposition \eqref{formula:global-local joint prob} of the joint probability, the update of each local factor is given by:
\begin{equation} \label{formula:local}
    \log q(\bm{\ell}_t)=\mathbb{E}_{q(\bm{g})}\!\left[\log p(\bm{y}_t,\bm{\ell}_t\mid \bm{g})\right]+\mathrm{const}.
\end{equation}
The update of the global factor is given by:
\begin{equation} \label{formula:global}
\log q(\bm{g})=
\log p(\bm{g})+
\sum_{t=1}^q
\mathbb{E}_{q(\bm{\ell}_t)}\!\left[\log p(\bm y_t,\bm{\ell}_t\mid \bm{g})\right]
+\mathrm{const}.
\end{equation}

Each update performs a coordinate-wise maximization of the ELBO with respect to a single block of variational parameters while keeping all remaining blocks fixed; see Algorithm~\ref{alg:global local CAVI}. Consequently, the ELBO sequence generated by the algorithm is monotonic non-decreasing. Since each local update is a conditional maximization of ELBO in a single coordinate direction, skipping a subset of local updates does not affect the monotonicity of the algorithm: blocks that are skipped remain unchanged, which is equivalent to performing a null update with zero improvement (see Supplementary Material A and~C). This observation is the foundation of our optimization scheme of CAVI to be introduced in the next section.


\begin{algorithm} [h]
\caption{\label{alg:global local CAVI} CAVI algorithm in the global-local framework}
\begin{algorithmic}[1] 
\Require Data $\bm y$; model $p(\bm y, \boldsymbol{\nu})$.

\State Initialize $\mathcal{L}(q) \gets -\infty$; global and local variational factors $q(\bm{g})$, $q (\bm{\ell}_t)$, $t = 1, \ldots, q$; iteration index $i = 0$.

\While{$\mathcal{L}(q)$ has not converged}
\State $i = i+1$
\For{$t = 1, \ldots, q$}
    \State Local update: update $q (\bm{\ell}_t)$ according to~\eqref{formula:local}
\EndFor
\State Global update: update $q(\bm{g}) $ according to~\eqref{formula:global}
\EndWhile
    
\State \Return $q(\bm{g})$, $q (\bm{\ell}_t)$, $t = 1, \ldots, q$. 

\end{algorithmic}
\end{algorithm}

\subsection{Adaptive focus scheme for CAVI } \label{sec:partial}

In sparse problems, the posterior assigns substantial mass to only a small subset of local units being ``active'', while the remaining units have posterior distributions concentrated near their inactive or null configuration. Updating of all local factors at every iteration (hereafter referred to as a ``full update'') may therefore be computationally inefficient, as many updates produce negligible changes in the variational objective. Since each local update can be interpreted as a coordinate-wise ascent step on the ELBO, it is natural to prioritize updates along directions that yield the largest ELBO gains. This motivates us to update only a subset of local factors that contribute mostly meaningfully to the ELBO in one iteration (hereafter referred to as a ``partial update'' ). 

To guide this selection, we introduce an \emph{activity score} based on posterior quantities available at CAVI iteration $i$. This score is model-dependent and reflects the current variational evidence that a given local factor contributes to explaining the data. Larger values indicate stronger posterior support for deviation from an inactive or null configuration.

In the example of hierarchical multi-response regression model with a spike-and-slab prior (Figure~\ref{fig:model}), the binary latent variables $\gamma_{st} \in \{0,1\}$ indicate whether predictor $s$ contributes to response $t$. Under a mean-field variational approximation that factorizes responses, the posterior inclusion probability (PPI) is approximated at iteration $i$ by the expectation under the current variational distribution $q^{(i)}$:
\begin{equation} \label{formula:ppi}
     p(\gamma_{st}=1 \mid \bm{y}) \;\approx\;
\mathbb{E}_{q^{(i)}}[\gamma_{st}],
\end{equation}
which is updated at each iteration and quantifies the current evidence for association. 
A natural measure of activity for unit $t$ is the variational probability, at iteration $i$, that at least one predictor is associated with unit $t$. Under the mean-field assumption, the indicators $\gamma_{1t}, \ldots, \gamma_{pt}$ are mutually independent under $q^{(i)}$, so this probability admits the closed form
\begin{equation}\label{formula:score_atlas}
a_t^{(i)} \;:=\; 
\mathrm{Pr}_{q^{(i)}}\!\left(\sum_{s=1}^p \gamma_{st} > 0 \right)
\;=\;
1 - \prod_{s=1}^p \left(1 - \mathbb{E}_{q^{(i)}}[\gamma_{st}]\right),
\end{equation}
which serves as a variational proxy for the posterior probability that unit $t$ is active. This quantity lies in $[0,1]$ and increases with the overall posterior support for activity of unit $t$.

At iteration $i$, we further define a binary selection vector
\[
\bm z^{(i)} = (z_1^{(i)}, \ldots, z_q^{(i)}) \in \{0,1\}^q,
\]
where $z_t^{(i)} = 1$ indicates that the local factor $t$ is selected for updating, generating the corresponding update set:
\[
\mathcal T^{(i)} \;\coloneqq\; \{ t \in \{1,\ldots,q\} : z_t^{(i)} = 1 \}.
\]
Only local factors with indices $t \in \mathcal T^{(i)}$ are updated at iteration $i$, while all remaining local factors are held fixed. Selection is performed by sampling independently
\[
z_t^{(i)} \sim \mathrm{Bernoulli}\!\left(\omega_t^{(i)}\right),
\]
with selection probabilities
\begin{equation} \label{formula:selection prob}
\omega_t^{(i)}
\;\coloneqq\;
\varepsilon^{(i)} + \bigl(1-\varepsilon^{(i)}\bigr)a_t^{(i)},
\end{equation}
where $\varepsilon^{(i)} \in [0,1]$ is a \emph{mixing parameter} controlling the balance between full updates and activity-guided partial updates. When $\varepsilon^{(i)}=1$, all local factors are updated, recovering standard CAVI. During the course of the algorithm, $\varepsilon^{(i)}$ decreases toward $0$, so that selection becomes increasingly driven by the activity scores, prioritizing factors with stronger posterior evidence of activity. This schedule encourages broad exploration during early iterations and progressively focuses computation on the subset of factors most relevant under the variational posterior. Examples of possible schedules and their impact on inference will be evaluated and discussed in Section~\ref{sec:compute}. 

In this way, selection of units is embedded directly within the CAVI procedure, leveraging the very nature of sparse modeling formulations. We hereafter refer to this procedure as the \emph{adaptive focus scheme for CAVI}, namely the AF-CAVI (Algorithm~\ref{alg:partial update}).

\begin{algorithm} [h]
\caption{\label{alg:partial update} AF-CAVI algorithm in the global-local framework}
\begin{algorithmic}[1] 
\Require Data $\bm y$; model $p(\bm y, \boldsymbol{\nu})$.

\State Initialize $\mathcal{L}(q) \gets -\infty$; global and local variational factors $q(\bm{g})$, $q (\bm{\ell}_t)$, $t = 1, \ldots, q$; iteration index $i = 0$;
\While{$\mathcal{L}(q)$ has not converged}

\State $i = i + 1$
\State Draw subset $\mathcal{T}^{(i)} \subset \{1, \ldots, q\}$ with adaptive selection probabilities $\bm{\omega}^{(i)} = (\omega^{(i)}_1, \dots \ \omega^{(i)}_q)$ defined in \eqref{formula:selection prob}.
\For{$t \in \mathcal{T} $}
    \State Update $q (\bm{\ell}_t)$ according to~\eqref{formula:local}
\EndFor
\State Update $q(\bm{g})$ according to~\eqref{formula:global}
\EndWhile

\State \Return $q(\bm{g})$, $q (\bm{\ell}_t)$, $t = 1, \ldots, q$. 
\end{algorithmic}
\end{algorithm}

\subsection{Related work} \label{sec:related work}

The optimization of mean-field CAVI under a global-local structure was first introduced by \textcite{hoffman_stochastic_2013} in the context of stochastic variational inference (SVI). In SVI, each local component corresponds to an individual observation and the ELBO decomposes into a sum of contributions over data points. Under conjugate exponential-family models, coordinate ascent updates for the variational factors admit closed-form expressions and these updates are equivalent to unit-length natural gradient steps on the ELBO with respect to the corresponding variational parameters. Moreover the natural gradient with respect to the global factor can also be written as a sum of per-observation contributions. This decomposition enables stochastic optimization by subsampling observations: at each iteration of SVI, only a minibatch of local factors is updated, yielding a stochastic but unbiased estimate of the natural gradient used to update the global factor. 

The SVI framework has been further extended to structured variational approximations allowing dependencies between global and local parameters \cite{hoffman_stochastic_2015}. However, these approaches still rely on exponential-family conjugacy and closed-form (or analytically tractable) natural-gradient expressions of the ELBO. To relax these requirements and extend stochastic variational methods to a broader range of complex models, \textcite{ranganath_black_2014} proposed a black-box variational inference (BBVI) method that approximates the gradient of the ELBO, when intractable, with black-box Monte Carlo estimators. The approximated gradient can then be employed within standard stochastic optimization procedures to maximize the ELBO. An example is the Quickdraws method \cite{loya_scalable_2025} developed for GWAS analysis at biobank scale, which considers spike-and-slab regression models where closed-form ELBO optimization is unavailable. In this case, Monte Carlo samples from the variational distribution provide unbiased gradient estimates, enabling scalable gradient-based optimization of the ELBO.

In contrast to gradient-based scalable VI methods such as SVI and BBVI, AF-CAVI does not require the computation of the gradient of the ELBO, either in closed form or via Monte Carlo estimation, and operates entirely within the coordinate ascent framework. Moreover, AF-CAVI does not rely on a specific hierarchical structure in which each local variational factor corresponds to an individual data point, as in classical SVI. Instead, local factors in AF-CAVI may be associated with arbitrary model-defined units, allowing the method to apply to a broader class of models. 

While related ideas of selective updating appear in other contexts, AF-CAVI differs fundamentally in both motivation and construction. Adaptive scanning strategies in MCMC \cite{richardson_bayesian_2011} and stratified subsampling schemes in SVI \cite{aliverti_stratified_2022} prioritize informative units, but operate within stochastic sampling frameworks. Randomized coordinate descent methods such as RACDM \cite{nesterov_efficiency_2012} adapt selection probabilities according to geometric properties of the objective (e.g., Lipschitz constants). In contrast, AF-CAVI embeds adaptive coordinate selection directly within deterministic variational inference, using posterior uncertainty as a principled, model-aware proxy for coordinate relevance. This integration of posterior-driven adaptivity into block coordinate ascent for BHMs constitutes a distinct methodological contribution.

\section{Results}

We next evaluate the performance of the AF-CAVI algorithm for BHMs in an application scenario of large-scale multi-trait pQTL mapping using the UK biobank proteomics data.  Section~\ref{sec:motivation}  introduces the biological motivation and dataset. Section~\ref{sec:atlasqtl} describes the BHM used for this problem and Section~\ref{sec:compute} presents the AF-CAVI implementation for this model. Section~\ref{sec:data_sim}  then systematically compares the computational efficiency and statistical performance of AF-CAVI against the standard CAVI under various simulation scenarios. Finally, Section~\ref{sec:real validate} reports genome-wide pQTL inference using the AF-CAVI–accelerated BHM and benchmarks it against univariate testing to quantify gains in power and scalability.



\subsection{A motivating case study: the genetic basis of the human proteome using the UK Biobank} \label{sec:motivation}


Understanding the genetic regulation of the human proteome is a key step towards linking genetic variation to disease mechanisms and potential therapeutic targets \cite{ferkingstad_large-scale_2021}. Nowadays, with the advancement of high-throughput proteomic technologies such Olink immuno assays \cite{assarsson_homogenous_2014}, 
it is possible to measure thousands of plasma protein assays across thousands of samples simultaneously, which opens the door to population-scale pQTL studies. Leveraging this technology, the UK Biobank Pharma Proteomics Project (UKB-PPP) provides measurements for nearly 3\,000 unique proteins of over 50\,000 UK Biobank participants \cite{sun_genetic_2022}. These participants were genotyped at about 850\,000 single nucleotide polymorphisms (SNPs), and data of over 90 million imputed variants were also made available \cite{bycroftUKBiobankResource2018}. This resource enables large-scale mapping of protein quantitative trait loci (pQTLs). The large sample size and high dimensionality of both genetic variants and protein traits provide a realistic and computationally demanding setting to demonstrate how AF-CAVI reduces the cost of applying BHMs to large-scale genome-wide studies. In this study, we use the UKB-PPP data of $n = 36\,626$ white British individuals (for population homogeneity) and their $q = 2\,919$ protein measurements after quality control. The sample, proteomic, and genetic data quality control filters applied are detailed in Supplementary Material B.

\subsection{The atlasQTL model for QTL hotspot detection} \label{sec:atlasqtl}

The scale of the UKB-PPP data enables not only the detection of individual pQTLs but also the study of shared regulatory structure across proteins. In particular, pQTL hotspots -- variants associated with multiple proteins -- offer a route to identifying pleiotropic regulatory mechanisms and shared pathways \cite{qian_large-scale_2022}. To detect such patterns, we consider the atlasQTL framework \cite{ruffieux_global-local_2020}, a multi-outcome hierarchical regression model that builds on the framework described in Section~\ref{sec:motivation_gl}. To target hotspot detection, the model introduces an additional hierarchical structure on $\gamma_{st}$, the indicator of association between SNP $s$ and trait $t$. Specifically, $\gamma_{st}$ follows a Bernoulli prior with success probability governed by a shared \emph{hotspot propensity} parameter $\theta_s$ that controls the probability of predictor $\bm{X}_s$ to be associated with multiple responses, and a response specific parameter $\zeta_t$ that adapts to the sparsity pattern corresponding to each response:
\[
\gamma_{st} \mid \theta_s, \zeta_t \sim \text{Bernoulli} \{\Phi (\theta_s + \zeta_t)\} ,
\] where $\Phi(\cdot)$ is the standard normal cumulative distribution function. The parameter $\theta_s$ follows a horseshoe prior \cite{carvalho_horseshoe_2010}, which shrinks most coefficients toward zero while retaining a small subset of large effects, consistent with the expectation that few SNPs are hotspots: 
\[
\theta_s \mid \lambda_s, \sigma_0 \sim \mathcal{N}(0, \lambda_s^2 \sigma_0^2), \quad  \lambda_s \overset{\mathrm{iid}}{\sim} \text{C}^{+}(0, 1), \quad \sigma_0 \sim \text{C}^+ (0, q^{-\frac{1}{2}}),
\] with $\text{C}^+(\cdot, \cdot)$ denoting the half-Cauchy distribution.
The parameter $\zeta_t$ follows a normal prior whose mean $n_0$ and variance $t_0^2$ are chosen to induce sparsity:
\[
\zeta_t \overset{\mathrm{iid}}{\sim} \mathcal{N}(n_0, t_0^2).
\]
In this way, information is shared across predictors and traits.

The main computational bottleneck of the atlasQTL CAVI algorithm lies in updating $\gamma_{st}$ and $\beta_{st}$ , which involves costly matrix multiplications and must be performed for all $s = 1, \dots, p$ and $t =1, \dots, q$. Restricting these updates to a small subset of indices at each iteration via adaptive focus should therefore effectively alleviate this burden (see Supplementary Material C for full details of the model, AF-CAVI updates, and hyperparameter specifications). The activity score defined for  subsampling local factors is defined in \eqref{formula:score_atlas}.

\subsection{AF-CAVI for atlasQTL} \label{sec:compute}

Since the variational objective is generally non-convex, coordinate ascent updates are not guaranteed to reach the global optimum and may depend strongly on initialization. To improve stability, we perform a short initialization phase in which all local factors are updated for a fixed number of iterations, denoted by $\text{I}_{init}$, before entering the adaptive-focus stage. From an optimization perspective, these full updates help to reduce sensitivity to initial variational parameters and allow subsequent iterations to focus on a subset of active components more reliably. In our experiments we set $\text{I}_{init}=15$. A sensitivity analysis for different $\text{I}_{init}$ values can be found in Supplementary Material~E. 


At iteration $i$, the mixing parameter $\varepsilon^{(i)}$ in \eqref{formula:selection prob} determines the extent to which updates are driven by uniform exploration versus activity scores. A decreasing schedule is required to transition from broad exploration to targeted updates. One option is to set $\varepsilon^{(i)}$ based on the ELBO increment between successive iterations (the \emph{ELBO scheme}), using it as a proxy for convergence speed. Given the large variability in ELBO differences, we stabilize this mapping via a logistic transformation,
\[
\varepsilon^{(i)}=\frac{1}{1+\exp\!\left[-\log\Delta\mathcal{L}^{(i)}(q)\right]},
\]
where $\Delta\mathcal{L}^{(i)}(q)=\mathcal{L}^{(i-1)}(q)-\mathcal{L}^{(i-2)}(q)$ for $i>2$ , (when $i=1$, $\varepsilon^{(i)} = 1$), AF-CAVI with this schedule is referred to as adaptive-focus with ELBO scheme (AFE) CAVI.

Although intuitive and free of additional hyperparameters, the ELBO scheme requires evaluating the ELBO at every iteration, which may introduce unnecessary computational cost in early stages. We therefore also consider an \emph{iteration scheme}, where $\varepsilon^{(i)}$ decreases deterministically at rate $\alpha \in (0, 1)$,
\[
\varepsilon^{(i)}=\alpha^{i-1}.
\]
AF-CAVI using this schedule is referred to as adaptive-focus with iteration scheme (AFI) CAVI. Empirical results show that performance is robust to the choice of $\alpha$ (Supplementary Material~F); throughout our experiments we use $\alpha=0.95$, which produces a smooth decay over roughly 500 iterations. This scheme removes the need for per-iteration ELBO evaluation. We further reduce overhead by monitoring convergence through intermittent ELBO evaluations, increasing the evaluation frequency as improvements diminish. AF-CAVI implemented with this strategy is referred to as adaptive-focus with \emph{optimized iteration scheme} (AFIO) CAVI.



\subsection{Simulations}\label{sec:data_sim}

\subsubsection{Data generation and simulation settings}
To emulate real settings whereby genotyping data entails local correlation patterns (linkage disequilibrium, LD), we use observed SNPs and simulate proteomic responses based on the data generation functions available from the R package \texttt{echoseq} \cite{ruffieux_global-local_2020}. The data generation procedure involves three steps. First, for a genomic region involving $p$ SNPs (\emph{block}), we select a percentage $a_p$ of ``active'' SNPs (i.e., SNPs associated with at least one protein) and a percentage $a_q$ of ``active'' proteomic responses (i.e., proteins associated with at least one SNP). We then define their association structure as follows. For each active SNP~$s$, we draw a \emph{pleiotropy parameter} from a $\text{Beta}(1, 5)$ distribution, i.e., the higher the pleiotropy of SNP $s$, the stronger its likelihood of being linked to multiple protein responses.  This setup generates variability in the number of associations per SNP, with a right-skewed distribution that favors sparse effects while allowing a few SNPs to act as \emph{hotspots}. 

Next, active responses are generated using a Gaussian linear additive model (i.e., each copy of the minor allele leads to a linear increase in risk).
The regression coefficients $\beta_{st}$ are parameterized through heritability components $h^2_{st}$, defined 
as the 
proportion of variance in response~$t$ explained by SNP~$s$. Here we first select an \emph{average total heritability level} $h^2_m$, and generate response-specific total heritability levels (i.e., variance in response~$t$ explained by all its associated SNPs) $h^2_t$, $t = 1, \ldots, q$, from $\text{Beta}\left(1, (1 - h^2_m)/h^2_m\right)$. Then, for each SNP $s = 1, \ldots, p$, we sample $h^2_{st}$ from a $\text{Beta}(2, 5)$ distribution, which ensures that most active SNPs have small effects, while a few have large effects. We then rescale the $h^2_{1t}, \ldots, h^2_{pt}$ so that $\sum_{s=1}^p h^2_{st} = h^2_t$.

Finally, to mimic realistic settings whereby the residuals exhibit correlation across traits (i.e., not accounted for by the available genetic data), we simulate the responses with noise that exhibits equicorrelation $\rho \in (0, 0.5)$ within groups of $50$ traits. Details of the simulation procedure are given in Supplementary Material~D.

The computational efficiency of the different CAVI implementations outlined in Section~\ref{sec:compute} is evaluated using relative performance metrics, defined as the percentage change in a given metric when comparing a given approach to the baseline ``Vanilla CAVI'', i.e., the CAVI algorithm implemented in the original atlasQTL algorithm \cite{ruffieux_global-local_2020}. These metrics are formally expressed as:
\begin{equation} \label{formula:metric}
    \Delta\text{metric} \ \text{(\%)} = \frac{\text{metric}_{\text{Method}} - \text{metric}_{\text{Vanilla}}}{\text{metric}_{\text{Vanilla}}} \times 100,
\end{equation}
where ``Method'' refers to the AFE, AFI or AFIO implementation. 
To evaluate computational efficiency, we focus on the \emph{relative runtime reduction}, denoted by $-\Delta\text{runtime}$, and the \emph{relative change in the total number of iterations}, $\Delta\text{iterations}$. The negative sign in $-\Delta\text{runtime}$ is introduced so that positive values correspond to shorter runtime compared to Vanilla CAVI, thereby providing a more intuitive interpretation of efficiency gains. Whereas for $\Delta\text{iterations}$, a positive value indicates the method requires more iterations to reach convergence and thus lower computational efficiency. 

Statistical performance 
is evaluated via precision, recall and false positive rate (FPR). These metrics capture the ability to recover true genetic signals while controlling spurious findings, which is critical in the context of sparse genetic association problems. 

\subsubsection{The benefits of multi-response BHM for pQTL mapping 
} \label{sec:toy}

\begin{figure}[h]
    \centering
    \includegraphics[width=\linewidth]{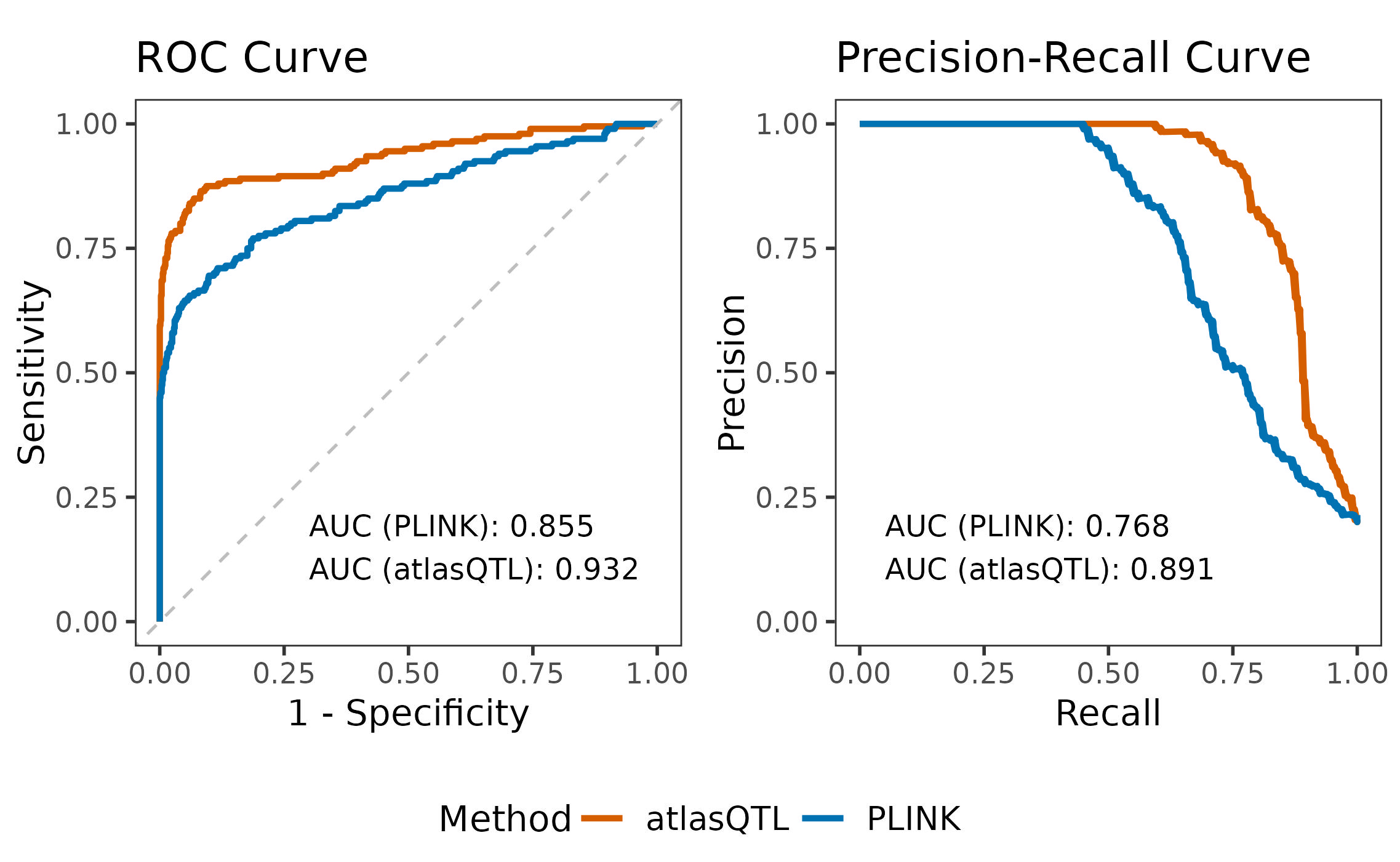}
  \caption{\textbf{Comparison of ROC (left) and PR (right) curves for atlasQTL and univariate testing in a toy example.} Scenario with many weak association signals emulating the genetic basis of complex traits: here a single active SNP (out of $200$) is associated with $200$ proteins (out of $1000$) and the average trait heritability for the active proteins is $0.05\%$.}
  \label{fig:toy}
\end{figure}

Before comparing the performance of AF-CAVI to the Vanilla version of CAVI, we present a toy simulation that demonstrates the advantage of BHMs for molecular QTL mapping, over conventional univariate testing. To limit the computational burden under the Vanilla algorithm, we choose a relatively small dataset, with $p = 200$ consecutive SNPs on chromosome 1 for all $n = 36\,626$ individuals. We then simulate associations for a single active SNP with $200$ responses out of simulated $q = 1\,000$ responses; this SNP therefore constitutes a hotspot. We also set a very low percentage of variance explained by the single SNP of $h^2_m = 0.05\%$ to simulate a challenging scenario with a large number of weak association signals. We then compare the statistical performance of univariate screening (using the \texttt{glm} function in PLINK2) and atlasQTL with Vanilla CAVI. To mitigate method-specific discrepancies caused by the differences in handling proxy signals due to the LD structure, we treat the $200$ SNPs as a single locus and summarize the output of each method by taking the minimum \emph{p}-value across all SNPs for each response in the univariate analysis, and the maximum posterior probability of inclusion (PPI) in the atlasQTL analysis. The corresponding Receiver Operating Characteristic (ROC) and Precision Recall (PR) curves shown in Figure~\ref{fig:toy} indicate that atlasQTL largely outperforms univariate screening, with AUROC of 0.932 vs 0.855, and AUPRC of 0.891 vs 0.768. This demonstrates the increased power of multivariate Bayesian methods which share information across co-regulated proteins to enhance the detection of weak pQTL signals. This observation has been extensively validated by \textcite{ruffieux_global-local_2020} in their evaluation of the atlasQTL model, as well as in other Bayesian QTL mapping contexts \cite{scott-boyer_integrated_2012, ruffieux_efficient_2017, kemper_multi-trait_2018}. 

\subsubsection{
Computational and statistical performance with AF-CAVI} \label{sec:sim validate}

We next evaluate the computational and statistical performance of AF-CAVI against Vanilla CAVI at UK Biobank scale. We simulate data matching the proteomics dataset in sample size ($n = 36\,626$) and number of proteins ($q = 3\,000$) across $p = 1\,000$ consecutive SNPs on chromosome 1. According to the Framingham Heart Study \cite{huan_systematic_2015} ($n =  5\,626$), the average heritability of the whole blood transcriptome is $0.13$ with 10\% display heritability $> 0.2$. Therefore, we use a reference simulation scenario with average heritability of $h^2_m = 0.15$, which we complete with two additional scenarios with lower ($h^2_m = 0.05$) and higher ($h^2_m = 0.3$) heritability levels. Drawing on the landmark univariate pQTL study of the UK Biobank by \cite{sun_plasma_2023}, which found that most active loci ($62.1\%$) are associated with a single protein ( $5^{\text{th}}-95^{\text{th}}$ quantiles of the distribution of the number of proteins associated with active loci are $1$ and $4$, respectively, with the median equal to 1), we set $a_q = 0.005$ as the general sparse scenario (leading to $15$ active proteins). We also simulate denser scenarios, namely with of $a_q = 0.2$ (leading to $600$ active proteins) and an extreme case of $a_q = 0.5$. Finally, we use $a_p = 0.01$ in all simulation scenarios (leading to $10$ active SNPs in total). We use $50$ replicates for each scenario. 

As outlined in Section~\ref{sec:compute}, we assess the performance of three different implementations of AF-CAVI against the Vanilla algorithm. We also further add a \emph{Randomly Focused} scheme of CAVI (RF-CAVI) that randomly updates $50\%$ of the local factors at each iteration, as a non-adaptive baseline for comparison with AF-CAVI's adaptive selection scheme (Table~\ref{tab:cavi_schemes}).

\begin{table}[ht]
\centering
\small
\begin{tabular}{ll}
\toprule
Implementation of CAVI& Coordinate ascent updating strategy\\
\midrule
Vanilla & Always update all local factors at each iteration\\
RF      & Randomly updating 50\% of local factors at each iteration\\
AFE     & AF-CAVI with mixing parameter $\varepsilon^{(i)}$ decreasing as a function of the ELBO \\
AFI     & AF-CAVI with mixing parameter $\varepsilon^{(i)}$ decreasing as a function of iteration $i$ \\
AFIO    & AFI optimized by not evaluating the ELBO at each iteration \\
\bottomrule
\end{tabular}
\caption{\textbf{Summary of different implementations of CAVI}.}
\label{tab:cavi_schemes}
\end{table}

\begin{figure}[!htbp]
  \centering
    \centering
    \includegraphics[width=\linewidth]{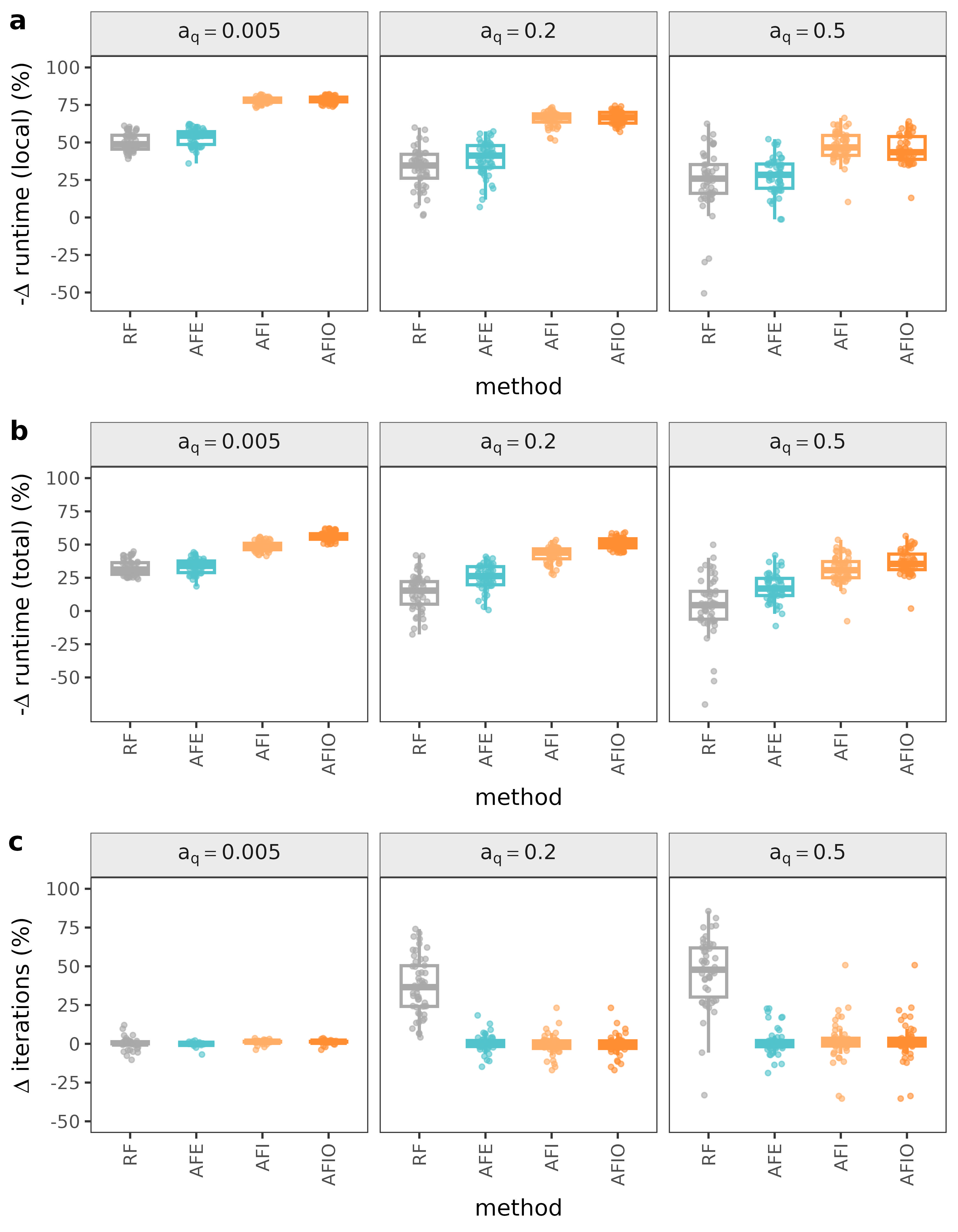}
    \caption{\label{fig:res_1}\textbf{Relative differences in runtime (local and total) and iterations for the RF-CAVI and series of AF-CAVI algorithms, with respect to the Vanilla CAVI for problems with various sparsity levels.} Problems are with average heritability $h_m^2 = 0.15$ and sparsity $a_q\in \{0.005, 0.2, 0.5\}$ (columns). Here we simulate 50 independent datasets with each $a_q$. The relative performance metrics employed are defined in~\eqref{formula:metric}. Relative differences in runtime are shown See Supplementary Material~G for results in simulation scenarios of different $h_m^2$ levels.}
\end{figure}

Figure~\ref{fig:res_1} compares the runtime reduction of  AF-CAVI schemes with RF and Vanilla CAVI in the reference simulation scenario of $h_m^2 = 0.15$. Overall, the adaptive focus methods significantly outperform RF-CAVI achieving greater runtime reduction with a very marginal increase in the number of iterations. The different AF-CAVI implementations show little difference in precision, recall and FPR, whereas RF-CAVI shows slightly lower precision and higher FPR (see Table~\ref{tab:res_sim} for results under $h^2_m = 0.15$ and $a_q = 0.005$, and Supplementary Material~G for the other simulation scenarios). To better understand where the computational gains arise, we decompose the runtime reduction by update type: in Vanilla CAVI, updating local factors accounts for $74.5\%$ of the per-iteration runtime, while updating global factors and evaluating the ELBO accounts for $15.2\%$ and $10.3\%$, respectively. We therefore report both the overall runtime reduction, $-\Delta\text{runtime (total)}$, and the reduction attributable specifically to local factor updates, $-\Delta\text{runtime (local)}$. A larger reduction in local computation time indicates that the adaptive focus strategy effectively restricts updates to a parsimonious subset of local factors.

Figure~\ref{fig:res_1}  also evaluates the efficiency of AF-CAVI in problems with different sparsity degrees, i.e., across different $a_q$ levels. It shows that the adaptive focus strategy has higher potential in runtime reduction for problems with higher sparsity (i.e., lower $a_q$). This finding is consistent across different heritability levels (Supplementary Material~G), and aligns with the design of AF-CAVI: when the sparsity level is high, the adaptive-focus strategy selects fewer local factors to update and thus achieves higher runtime reduction. Although the overall potential for runtime reduction naturally diminishes for both RF and AF-CAVI as sparsity decreases, the advantage of AF-CAVI in these less sparse scenarios is still obvious. In scenarios with $a_q = 0.2$ and $0.5$, both RF and AF-CAVI exhibit increased variability in the number of iterations. Nevertheless, AF-CAVI maintains better control over iteration increase compared to RF-CAVI, with the latter showing increases of up to approximately 25\%. This suggests that when a larger fraction of local factors are truly active, convergence toward the optimal ELBO depends on consistently updating a sufficiently large set of informative components, a requirement that random selection does not reliably satisfy. 

Among all the adaptive focus schemes, AFIO-CAVI demonstrates the highest reduction in runtime, achieving about $75\%$ reduction in local factor runtime and $50\%$ in total runtime in the general sparse scenario ($a_q = 0.005$). While the best-performing AF-CAVI scheme may vary across different models and applications, the consistent patterns observed here suggest that the advantage of adaptive focus schemes is broadly generalizable.  Hereafter, we will use our best-performing scheme, the AFIO-CAVI, for the rest of the paper. The algorithm is implemented in a stand-alone R package called \texttt{AFatlasQTL}.

\begin{table}[!htbp]
\centering
\begin{tabular}{lccc}
\toprule
Method & Precision & Recall & FPR ($\times 10^{-5}$) \\
\midrule
Vanilla
& $0.73 \pm 0.02$
& $0.92 \pm 0.01$
& $0.36 \pm 0.03$ \\

RF
& $0.70 \pm 0.02$
& $0.92 \pm 0.01$
& $0.39 \pm 0.03$ \\

AFE
& $0.73 \pm 0.02$
& $0.92 \pm 0.01$
& $0.36 \pm 0.03$ \\

AFI
& $0.73 \pm 0.02$
& $0.92 \pm 0.01$
& $0.36 \pm 0.03$ \\

AFIO
& $0.73 \pm 0.02$
& $0.92 \pm 0.01$
& $0.36 \pm 0.03$ \\
\bottomrule
\end{tabular}
    \captionof{table}{\label{tab:res_sim} \textbf{Comparison of precision, recall and FPR of the different CAVI implementations in the reference simulation scenario} ($h_m^2 = 0.15, a_q = 0.005$). PPI threshold of 0.5 is used to report active signals, following to the \emph{median probability model rule} of \cite{barbieri_optimal_2004}. Average metrics are reported with ± standard error computed across 50 replicates. FPR values are reported in units of $10^{-5}$. }
\end{table}

\subsection{Efficient multi-trait pQTL mapping 
using the UK Biobank plasma proteomics data 
}\label{sec:real validate}

Having established the computational and statistical properties of AF-CAVI in controlled simulations, we now assess its practical utility in a real-data setting by conducting genome-wide multi-trait pQTL mapping in the UK Biobank plasma proteomics data.

To enable genome-wide multi-trait pQTL inference with the AF-CAVI–accelerated BHM, we design a pipeline that partitions each chromosome into approximately LD-independent blocks, allowing parallel computation, and jointly models multiple proteomic traits within each block (Figure~\ref{fig:pipeline}). To mitigate any irrelevant differences arising from proxy signals due to LD, we summarize the results \emph{locus-wise} by merging overlapping signals.  Specifically, starting from SNPs with the largest number of associated proteins (or highest maximum PPI by SNP when there are ties of numbers of proteins associated), we group together SNPs within $\pm 0.5$ Mb that share associated proteins, treating them as a single locus rather than distinct signals.  For each locus, we define the \emph{sentinel SNP} as the SNP with largest number of associated proteins (or highest maximum PPI by SNP when there are ties of numbers of proteins associated), and refer to the corresponding pQTL signals as \emph{sentinel associations}.

We illustrate this pipeline by applying atlasQTL with AF-CAVI to chromosome 1. After quality control, the chromosome is partitioned into $124$ LD blocks comprising $160\,702$ SNPs in total, with an average of $1\,296$ SNPs per block. Block boundaries are defined using LDetect \cite{berisa_approximately_2016} based on the 1000 Genomes Phase 1 reference panel. Within each block, $2\,919$ proteomic traits are modelled jointly. To capture a range of settings, we select six representative blocks differing in size and sparsity. On these blocks, we compare AF-CAVI with Vanilla CAVI in terms of computational efficiency, and with univariate testing in terms of pQTL discovery.

\begin{figure}[ht]
\centering
\small
\begin{tikzpicture}[node distance=1.5cm]

\node (start) [input] {
  \setstretch{1}
  Genotyping data
};

\node (ldblock) [greenblock, below=0.5cm of start] {
  \setstretch{1}
  Divide each chromsome into LD-independent blocks
};

\node (proteomics) [input, right=0.5cm of ldblock] {
  \setstretch{1}
  Proteomics data\\
  (multi-trait)
};

\node (bayes) [redblock, below=0.5cm of ldblock, xshift=2.1cm] {
  \setstretch{1}
  Joint pQTL mapping with all proteomic responses
  (atlasQTL with AF-CAVI) 
};

\node (summarise) [yellowblock, below=0.5cm of bayes] {
  \setstretch{1}
  Locus-wise summarization of pQTL signals
};

\node (output) [output, below= 0.5cm of summarise] {
  \setstretch{1}
  List of loci with associated 
  proteomic responses
};

\draw [arrow] (start) -- (ldblock);
\draw [arrow] (ldblock) -- (bayes);
\draw [arrow] (proteomics) -- (bayes);
\draw [arrow] (bayes) -- (summarise);
\draw [arrow] (summarise) -- (output);

\end{tikzpicture}
\caption{ \textbf{Pipeline for genome-wide joint pQTL mapping using AF-CAVI-accelerated BHM \cite{ruffieux_global-local_2020}.} }
\label{fig:pipeline}
\end{figure}

Table~\ref{tab:real} reports the computational performance of AF-CAVI (AFIO implementation with 15 initial full updates and $\alpha = 0.95$) compared to Vanilla CAVI. On an Intel Xeon Platinum 8276 2.2 GHz machine, the AF-CAVI reduces runtime by approximately $50\%$, which is in line with the simulation results of Section~\ref{sec:sim validate}. Peak memory usage remains below 5 GB per block, demonstrating that the approach is both time- and memory-efficient under parallel execution. 

Table~\ref{tab:real2} further compares the pQTL signals identified by AF-CAVI accelerated BHM with univariate testing. We use the Bonferroni threshold \emph{p}-value < $1.7 \times 10^{-11}$ ($5 \times 10^{-8}$ divided by $2\,919$, the number of proteins tested) for univariate testing, and a stringent threshold of  $\text{PPI} > 0.99$ for BHM. BHM not only reports more active sentinel SNPs, but also more sentinel pQTL associations and larger hotspots. Together, these results demonstrate that AF-CAVI enables scalable genome-wide multi-trait pQTL inference while substantially improving signal discovery over univariate testing, combining  computational efficiency and enhanced statistical power in biobank-scale data. 
\begin{figure*}[!htbp]
\centering

\begin{minipage}{\textwidth}
\centering
\scriptsize
\begin{tabular}{c|rrr|cc|cc|cc}

\toprule
\textbf{block ID} & \textbf{block start (bp)} & \textbf{block end (bp)} & \textbf{p} 
& \multicolumn{2}{c|}{\textbf{Runtime (min)}} 
& \multicolumn{2}{c|}{\textbf{Iterations}} 
& \multicolumn{1}{c}{\textbf{Memory (GB)} } \\
 & & & & Vanilla CAVI & AF-CAVI & Vanilla CAVI & AF-CAVI I \\
\midrule
1 & 10583 & 1892607 & 785 & 8.23 & 5.32 & 197 & 200 & 3.65 \\
21 & 34799758 & 37549183 & 1690 & 46.85 & 22.68 & 486 & 489 & 3.78 \\
65 & 113273306 & 114873845 & 1147 & 20.04 & 10.60 & 328 & 330 & 3.57 \\
67 & 115880593 & 118839067 & 2110 & 98.26 & 41.69 & 584 & 588 & 4.29 \\
73 & 154770403 & 156336133 & 828 & 9.05 & 5.80 & 216 & 218 & 3.51 \\
85 & 178944309 & 181144121 & 1401 & 55.92 & 15.89 & 396 & 398 & 3.81 \\
\bottomrule
\end{tabular}
\captionof{table}{\label{tab:real} \textbf{Comparison of runtime, number of iterations and peak running memory usage between the Vanilla CAVI and AF-CAVI on selected blocks of chromosome 1.} The blocks are selected to represent various sparsity levels and numbers of SNPs across chromosome 1. The peak memory usage of both methods is identical and thus only one column is shown. To accommodate for different lengths of genomic blocks, we use a prior proportional to the number of SNPs (See Supplementary Material~C). The AFIO implementation for AF-CAVI is used with $\alpha = 0.95$, $\text{I}_{\text{init}} = 15$.}
\end{minipage}

\vspace{1em} 

\begin{minipage}{\textwidth}
  \centering
  \scriptsize
\begin{tabular}[t]{ccccc}
\toprule
 & \multicolumn{2}{c}{\textbf{\# Sentinel SNPs}} & \multicolumn{2}{c}{\textbf{\# Sentinel pQTL associations}} \\
\cmidrule(lr){2-3} \cmidrule(lr){4-5}
block id & Univariate & BHM & Univariate & BHM \\
\midrule
1 & 3 & 8 & 3 & 8\\
21 & 2 & 2 & 2 & 14\\
65 & 2 & 3 & 7 & 18\\
67 & 3 & 14 & 3 & 19\\
73 & 2 & 18 & 11 & 35\\
85 & 1 & 6 & 1 & 122\\
\bottomrule
\end{tabular}
  \captionof{table}{\label{tab:real2} \textbf{Comparison of the number of sentinel SNPs and associations reported by Bayesian hierarchical modeling (BHM) and univariate testing (Univariate) by block.} The threshold for reporting active pQTL signals is PPI $>0.99$ and \emph{p}-value < $1.7 \times 10^{-11}$ for BHM and univariate testing respectively.  Loci are defined by an iterative merging procedure: starting from the SNP with highest number of associated proteins (or with highest PPI / lowest \emph{p}-value when there are ties of association proteins), within a distance of $\pm 0.5$ Mb, any SNPs associated with overlapping proteins are collapsed into a single locus rather than treated as independent signals. The ``sentinel SNP'' is defined as the SNP with highest number of associated proteins in the locus (or with highest PPI / lowest \emph{p}-value when there are ties of association proteins). ``Sentinel associations'' are pQTL signals associated with the sentinel SNP in the locus.  }
\end{minipage} 

\end{figure*}

To illustrate how the additional signals detected by BHM inform biological interpretation we consider the pQTL hotspot at SNP rs2476601 as a case study. This variant is located within the protein tyrosine phosphatase non-receptor type 22 (\emph{PTPN22}) gene on chromosome 1 and corresponds to a nonsynonymous coding variant that results in an arginine-to-tryptophan substitution at position 620 (R620W) of the PTPN22 protein \cite{siminovitch_ptpn22_2004}. \emph{PTPN22} encodes a tyrosine phosphatase that negatively regulates T-cell receptor signalling 
\cite{bottini_functional_2004}. The R620W polymorphism is a well-established risk factor for multiple autoimmune diseases, including rheumatoid arthritis, inflammatory bowel disease, type~1 diabetes, systemic lupus erythematosus and autoimmune thyroid disease, particularly in cohorts of European ancestry \cite{begovich_missense_2004,bottini_functional_2004,kyogoku_genetic_2004,heward_association_2007,lee_ptpn22_2007,totaro_ptpn22_2011,spalinger_loss_2021,xue_genetic_2013,de_jager_evaluating_2006,hedjoudje_rs2476601_2017}.

Both BHM and univariate testing report multiple proteins associated with rs2476601. However, BHM detects 10 additional associations beyond the 6 identified by univariate analysis. 
All 16 pQTL signals are classified as \emph{trans}, defined here as SNPs located more than $1$ Mb away from the transcription start site (TSS) of the gene encoding the associated protein or on a different chromosome. In our case, with the exception of IL10 (encoded on chromosome 1, but more than $1$ Mb from rs2476601), all associated proteins are encoded on chromosomes other than chromosome 1 (Supplementary Material~H). 

To characterize the functional context of these associations, we perform over-representation analysis (ORA) on the set of associated proteins. ORA evaluates whether the observed set of proteins is enriched in predefined pathways relative to a background universe using Fisher’s exact test on a $2 \times 2$ contingency table \cite{agapito_comprehensive_2021, marco-ramell_evaluation_2018}. We obtain pathway annotations from the KEGG database \cite{ogata_computation_1998}, and compute enrichment using the \texttt{enrichKEGG} function from the \texttt{clusterProfiler} R package. We assess statistical significance using one-sided Fisher’s exact test $p$-values adjusted by the Benjamini--Hochberg procedure with FDR threshold of 0.05. 

Using the 16 proteins identified by BHM, we detect 20 enriched pathways (Table~\ref{tab:ora}), whereas univariate testing yields only a single enriched pathway. We note that ORA results based on small input sets should be interpreted with caution: with only 14 proteins mapped to KEGG, single-protein changes can substantially alter the set of enriched pathways and many of the immune-related categories share overlapping member genes. The pathways reported here are therefore best viewed as hypothesis-generating rather than definitive functional annotations. Several of the pathways identified by BHM are directly related to immune-mediated diseases, including inflammatory bowel disease \cite{spalinger_loss_2021} and allograft rejection \cite{sfar_ptpn22_2009}, both previously linked to the R620W variant. More broadly, enrichment of immune and infectious disease pathways is consistent with downstream effects of altered immune activation mediated by \emph{PTPN22}.

Taken together, the additional associations identified by BHM broaden the set of candidate proteins linked to rs2476601 and the corresponding ORA highlights immune signalling categories consistent with the known role of \emph{PTPN22} in immune regulation. While these candidate associations require independent replication and orthogonal evidence (for example, colocalization or conditional analyses) before mechanistic claims can be made, they illustrate how the increased sensitivity of joint hierarchical modeling can generate testable hypotheses for follow-up functional studies.

\begin{table}[!htbp]
    \centering
    \scriptsize
\begin{tabular}[t]{llp{4.5cm}llll}
\toprule
ID & Category & Description & Gene Ratio & Background Ratio & Adj. \emph{p}-value\\
\midrule
hsa04620 & Immune system & Toll-like receptor signaling pathway & 5/14 & 41/1722 & 0.00\\
hsa05146 & Infectious disease: parasitic & Amoebiasis & 4/14 & 45/1722 & 0.01\\
hsa05330 & Immune disease & Allograft rejection & 3/14 & 19/1722 & 0.01\\
hsa04622 & Immune system & RIG-I-like receptor signaling pathway & 3/14 & 20/1722 & 0.01\\
hsa05143 & Infectious disease: parasitic & African trypanosomiasis & 3/14 & 20/1722 & 0.01\\
hsa05140 & Infectious disease: parasitic & Leishmaniasis & 3/14 & 32/1722 & 0.02\\
hsa04061 & Signaling molecules and interaction & Viral protein interaction with cytokine and cytokine receptor & 4/14 & 71/1722 & 0.02\\
hsa04060 & Signaling molecules and interaction & Cytokine-cytokine receptor interaction & 6/14 & 188/1722 & 0.02\\
hsa05321 & Immune disease & Inflammatory bowel disease & 3/14 & 36/1722 & 0.02\\
hsa04625 & Immune system & C-type lectin receptor signaling pathway & 3/14 & 38/1722 & 0.02\\
hsa05133 & Infectious disease: bacterial & Pertussis & 3/14 & 38/1722 & 0.02\\
hsa04512 & Signaling molecules and interaction & ECM-receptor interaction & 3/14 & 43/1722 & 0.02\\
hsa05145 & Infectious disease: parasitic & Toxoplasmosis & 3/14 & 44/1722 & 0.02\\
hsa05142 & Infectious disease: parasitic & Chagas disease & 3/14 & 47/1722 & 0.03\\
hsa05164 & Infectious disease: viral & Influenza A & 3/14 & 54/1722 & 0.04\\
hsa05152 & Infectious disease: bacterial & Tuberculosis & 3/14 & 61/1722 & 0.04\\
hsa05171 & Infectious disease: viral & Coronavirus disease - COVID-19 & 3/14 & 61/1722 & 0.04\\
hsa04062 & Immune system & Chemokine signaling pathway & 3/14 & 63/1722 & 0.04\\
hsa04940 & Endocrine and metabolic disease & Type I diabetes mellitus & 2/14 & 22/1722 & 0.04\\
hsa05134 & Infectious disease: bacterial & Legionellosis & 2/14 & 24/1722 & 0.05\\
\bottomrule
\end{tabular}
    \caption{\textbf{KEGG ORA enrichment results for proteins associated with rs2476601 reported by BHM.} Columns report the KEGG pathway identifier (ID), high-level functional category categories (Category), pathway description (Description), the proportion of associated proteins annotated to each pathway (Gene Ratio), the corresponding proportion in the entire set of 2919 tested proteins (Background Ratio), and the Benjamini–Hochberg–adjusted one-sided Fisher’s exact test \emph{p}-value (Adj. \emph{p}-value). Note that only $1\,722$ out of the total $2\,919$ tested proteins and 14 out of the 16 reported active proteins associated with rs2476601 are included in the KEGG database, and thus the gene and background ratios are divided by 14 (instead of 16) and $1\,722$ (instead of $2\,919$).}
    \label{tab:ora}
\end{table}

\section{Discussion} \label{sec: conclusion}

Motivated by the computational demands in biobank-scale analysis, we have introduced an \emph{adaptive focus strategy} for VI, AF-CAVI, tailored to large Bayesian hierarchical models. AF-CAVI selectively updates local variational factors corresponding to units whose activity is supported by the data as the algorithm progresses. This results in a probabilistic framework based on dynamically updated \emph{activity scores}, designed to balance computational efficiency with stable convergence and adequate exploration of the parameter space. 

We evaluated the extent to which this adaptive strategy reduces computational cost without degrading statistical performance in the context of multi-trait pQTL mapping using UKB-PPP data. In simulated settings emulating the genetic basis of $3\,000$ proteins, (which corresponds to the current scale of the UKB-PPP datasets), our best-performing AF-CAVI scheme achieves up to 75\% runtime reduction in local update time and around 50\% reduction in total time under sparse regimes, while maintaining precision, recall and FPR virtually unchanged compared to the Vanilla CAVI algorithm. Comparisons with RF-CAVI, which updates random subsets of local factors, further highlighted that adaptivity is critical: AF-CAVI reduces per-iteration cost while controlling the number of iterations needed for convergence, whereas random sub-sampling leads to a less favorable trade-off.

These findings are supported in the real data analysis of the UKB-PPP data. By partitioning the genome into approximately LD-independent blocks, multi-trait pQTL mapping can be executed in parallel across computational units, making genome-wide analysis tractable. In this setting, AF-CAVI retains the computational gains observed in simulations. Moreover, comparisons with univariate testing indicate that joint hierarchical modeling increases the number and biological interpretability of detected signals, suggesting improved power in large-scale proteomic studies.

The computational gains achieved by AF-CAVI will vary according to the model and data at hand, particularly depending on the level of sparsity. The atlasQTL multi-trait pQTL mapping example is a general family of hierarchically linked sparse regressions, which are widely applied for association estimation. Beyond this setting, AF-CAVI is also well suited to models with spike-and-slab priors, such as network models for gene co-expression, where sparsity in edge structure can be exploited to reduce computational cost.

In summary, AF-CAVI bridges the gap between the flexibility of Bayesian joint modeling for sparse problems and the computational demands inherent in large-scale, high-dimensional real-world analyses. By selectively allocating computation to data-supported components, it retains the flexibility of hierarchical models while mitigating their computational burden. The framework is broadly applicable and can be extended to other structured inference problems, offering a principled route to applying Bayesian methods in large-scale genomic and biobank analyses.


\section{List of abbreviations}

\begin{itemize}
    \item \textbf{GWAS}: Genome-wide association studies
    \item \textbf{QTL}: Quantitative trait locus
    \item \textbf{pQTL:} Protein quantitative trait locus 
    \item \textbf{BHM}: Bayesian hierarchical models
    \item \textbf{MCMC:} Markov Chain Monte Carlo
    \item \textbf{VI:} Variational inference
    \item \textbf{CAVI:} Coordinate ascent variational inference
    \item \textbf{AF-CAVI}: Adaptive-focus CAVI
    \item \textbf{UKB-PPP:} the UK Biobank Pharma Proteomics Project
\end{itemize}

\section{Software availability}

AF-CAVI for atlasQTL is implemented in an open-source R package \texttt{AFatlasQTL}, available at \url{https://github.com/yiran2000/AFatlasqtl}. 

\section{Funding information} 
This work was funded by UKRI Programme Grant MC\_UU\_00040/01, and supported by the Lopez–Loreta Foundation (H.R.). For the purpose of open access, the authors have applied a Creative Commons Attribution (CC BY) license to any Author Accepted Manuscript version arising.

\section{Acknowledgments}

We thank Colin Starr for his valuable assistance with the algorithmic and R package development, as well as Prof. Adam Butterworth and Dr. Paul Lyons for their helpful insights and discussions.

\newpage
\onecolumn
\nolinenumbers
\printbibliography

\newpage
\appendix

\appendix
{\Huge \bfseries Supplementary Material  \par}
\vspace{1em} 

\section{Monotonic increase of the ELBO after each local and global update}
Recall the definition of ELBO in Section 2.3:
\begin{equation} \label{formula:elbo}
\begin{aligned}
    \mathcal{L}(q(\bm{\nu}))
    &= \mathbb{E}_{q(\bm{\nu})}[\log p(\bm{y}, \bm{\nu})] - \mathbb{E}_{q(\bm{\nu})}[\log q(\bm{\nu})].
\end{aligned}
\end{equation}
With the definition of KL divergence, ELBO and mean-field factorization, the ELBO can be decomposed as:
\begin{equation} \label{formula:elbo decomp}
\begin{aligned}
    \mathcal{L}(q(\bm{\nu})) 
    &= \mathbb{E}_{q(\bm{g})}[\log p(\bm{g}) - \log q(\bm{g})] + \sum_t\mathbb{E}_{q(\bm{\ell}_t)}[\mathbb{E}_{q(\bm{g})}\log p(\bm{\ell}_t, \bm{y}_t|\bm{g}) - \log q(\bm{\ell}_t)]. \\
\end{aligned}
\end{equation}
To evaluate the change of ELBO after the update of local factor~$t$ in the form of 
\begin{equation}\label{formula:local_app}
    \log q^{\mathrm{new}}(\bm{\ell}_t)=\mathbb{E}_{q(\bm{g})}\!\left[\log p(\bm{y}_t,\bm{\ell}_t\mid \bm{g})\right]+\mathrm{const},
\end{equation}
re-write the ELBO in (\ref{formula:elbo decomp}) as a part involving factor $t$ and a constant part that do not depend on factor $t$ :
\begin{equation} \nonumber
\begin{aligned}
\mathcal{L}(q(\bm{\nu})) 
&= \mathbb{E}_{q(\bm{\ell}_t)}[\mathbb{E}_{q(\bm{g})} [\log p(\bm{\ell}_t, \bm{y}_t \mid \bm{g})] - \log q(\bm{\ell}_t)] +\mathrm{const} \\
&= -\text{KL} \left[q(\bm{\ell}_t) \,\Vert\, \tilde p_t(\bm{\ell}_t) \right] +\mathrm{const}.
\end{aligned}
\end{equation}
where\[\quad
\tilde p_t(\bm{\ell}_t)
\propto
\exp\!\left\{
\mathbb{E}_{q(\bm{g})}\left[\log p(\bm{\ell}_t,\bm{y}_t \mid \bm{g})\right]
\right\}.
\]
The CAVI update in~\eqref{formula:local_app} sets 
$q^{\mathrm{new}}(\bm{\ell}_t) = \tilde p_t(\bm{\ell}_t)$, 
driving the KL term to zero. The resulting increase in the ELBO 
equals the previous value of 
$\text{KL}[q^{\mathrm{old}}(\bm{\ell}_t) \,\Vert\, 
\tilde p_t(\bm{\ell}_t)]$.
Similarly, with the global update 
\begin{equation} \nonumber
\log q^{\mathrm{new}}(\bm{g})=
\log p(\bm{g})+
\sum_{t=1}^q
\mathbb{E}_{q(\bm{\ell}_t)}\!\left[\log p(\bm y_t,\bm{\ell}_t\mid \bm{g})\right]
+\mathrm{const},
\end{equation}
re-write the ELBO as
\begin{equation} \nonumber
\begin{aligned}
\mathcal{L}(q(\bm{\nu})) 
&= \mathbb{E}_{q(\bm{g})} \left[ \log p(\bm{g}) + \sum_t \mathbb{E}_{q(\bm{\ell}_t)}[\log p(\bm{\ell}_t, \bm{y}_t|\bm{g})] - \log q(\bm{g}) \right] +\mathrm{const} \\
&= -\text{KL}  \left[
q(\bm{g}) \,\middle\|\, \tilde p(\bm{g})
\right] + \mathrm{const},
\end{aligned}
\end{equation}
where\[
\tilde p(\bm{g})
\propto
\exp\left\{
\log p(\bm{g}) 
+ \sum_t \mathbb{E}_{q(\bm{\ell}_t)}[\log p(\bm{\ell}_t, \bm{y}_t \mid \bm{g})]
\right\}.
\]
The ELBO increase from this update equals the value of 
$\text{KL}[q^{\mathrm{old}}(\bm{g}) \,\|\, \tilde p(\bm{g})]$ 
prior to the update.


\clearpage
\section{Quality control of the UKB-PPP data} \label{append:qc}
The currently available UKB-PPP dataset contains $2\,923$ unique proteins measured for $53\,021$ participants at their initial assessment visit from 2006 to 2010. Genome-wide genotyping data is available for $52\,567$ participants. Measurements of their repeated assessments and COVID-19 repeat imaging study are not included in our study. We further limit our analysis to the $43\,711$ samples self-identified as white British and have similar genetic ancestry based on a principal components analysis of the genotypes. We then removed 23 samples with likely sex swaps, and exclude 318 samples with duplicates or second-degree relations in the dataset (randomly excluded one sample from the 318 pairs of kinship coefficient (corresponding to second-degree or closer relatedness) over 0.1769 estimated by KING \cite{manichaikul_robust_2010}). Noticing that 15.5\% of the samples miss measurements on entire panels of cardiometabolic II, inflammation II, neurology II, and oncology II, we also remove such samples to avoid bias caused by missingness. Finally, we remove proteins with over 20\% of missingness in the rest of the dataset, leaving $q = 2\,919$ unique proteins and $n = 36\,626$ samples. All the remaining samples have genotyping rate over 99\%, and thus no further samples are removed.

Imputed genomic variants of UK Biobank are stored as ``gene dosages'', which are triplets containing the probability of the variant containing two copies of the reference allele (homozygous reference genotype), one copy of the reference allele and one copy of the alternative allele (heterozygous genotype), or two copies of the alternative allele (homozygous alternative genotype). We translate gene dosages to a 0-2 scale during the additive dose-effect scheme, i.e.,
\begin{equation}
    \hat{g}_{is} = \mathbb{E}(g_{is}) = \mathbb{P}(g_{is} = 0) \times 0 + \mathbb{P}(g_{is} = 1) \times 1 + \mathbb{P}(g_{is} = 2) \times 2 , \nonumber 
\end{equation} 
where $\hat{g}_{is}$ denotes the (estimated) genotype for the $s^{\text{th}}$ SNP of $i^{\text{th}}$ sample, and 0, 1, 2 refer to the homozygous reference genotype, heterozygous genotype, homozygous alternative genotype respectively. 

Within each block, we filter for high-quality imputed variants defined by: Info score > 0.8, MAF > 0.01 (in the entire UK Biobank cohort), MAC > 20 (in our QC-filtered samples), HWE exact test \emph{p}-value > $10^{-15}$, LD pruning with 1000 variant windows, 100 sliding windows and $r^2$ < 0.9, and > 95\% of the expected gene dosage within 0.1 of either 0, 1 or 2. We adjust the NPX values for age, $\text{age}^2$, sex, age $\times$ sex,  $\text{age}^2$ $\times$ sex, batch, UKB centre, UKB genetic array, time between blood sampling and measurement and the first 20 genetic PCs, and then conduct pQTL inference on the regression residuals.

\clearpage
\section{Details of the atlasQTL model and AF-CAVI updates}   
Here we provide details of the atlasQTL model by \textcite{ruffieux_global-local_2020}. Given $p$ candidate predictors $\bm{X} = (\bm{X}_1, ..., \bm{X}_p)$ and $q$ responses $\bm{y} = (\bm{y}_1, ..., \bm{y}_q)$, atlasQTL constructs a series of hierarchically related regressions:
\begin{equation} \nonumber
\begin{aligned}
& \bm{y}_t \mid \boldsymbol{\beta}_t, \tau_t \sim N_n(\bm{X} \boldsymbol{\beta}_t, \tau_t^{-1} \mathbf{I}_n),  \quad t = 1, ..., q ,  \\
& \beta_{st} \mid \gamma_{st}, \tau_t, \sigma \sim \gamma_{st} N(0, \sigma^2 \tau_t^{-1}) + (1 - \gamma_{st}) \delta_0, \quad s = 1, ..., p , \\
& \gamma_{st} \mid \theta_s, \zeta_t \sim \text{Bernoulli} \{\Phi (\theta_s + \zeta_t)\}, \quad\zeta_t \overset{\mathrm{iid}}{\sim} N(n_0, t_0^2) , \\
& \theta_s \mid \lambda_s, \sigma_0 \sim N(0, \lambda_s^2 \sigma_0^2), \quad  \lambda_s \overset{\mathrm{iid}}{\sim} C^{+}(0, 1), \quad \sigma_0 \sim C^+ (0, q^{-\frac{1}{2}}) ,
\end{aligned}
\end{equation}
where $\delta_0$ is the Dirac distribution, $\Phi(\cdot)$ is the standard normal cumulative distribution function, and $C^+(\cdot, \cdot)$ is half-Cauchy distribution. The key parameters in this model are: 
\begin{itemize}
    \item $\beta_{st}$: the regression coefficient between predictor $\bm{X}_s$ and response $\bm{y}_t$;
    \item $\gamma_{st}$: a binary latent variable that takes value 1 when there is an association between $\bm{X}_s$ and $\bm{y}_t$ and 0 otherwise;
    \item $\theta_s$: a hotspot propensity that controls the probability of predictor $\bm{X}_s$ to be associated with multiple responses;
    \item $\zeta_t$: response specific parameter that adapts to the sparsity pattern corresponding to each response.
\end{itemize}

The model facilitates a joint modeling between predictors and responses in the way that each response, $\bm{y}_t$, is related linearly to the predictors $\bm{X}$ with a specific precision $\tau_t$. The responses are conditionally independent across the regressions, while their dependence structure is captured via shared parameters $\sigma^2$ and $\theta_s$, which are common to all the responses. Besides, $\zeta_t$ is shared across predictors. This naturally serves co-selection of predictors and responses by leveraging strength across responses associated with the same predictors, as well as predictors associated with the same responses. 


The closed-form CAVI updates are provided by \textcite{ruffieux_global-local_2020}. Under the atlasQTL model structure, each unit corresponds to one regression with respect to response $\bm{y}_t$. Correspondingly, the global parameters shared across responses are $\bm{g} = (\sigma, \sigma_0, \lambda_1, \dots, \lambda_p, \theta_1, \dots, \theta_p)$, and local parameters $\bm{\ell}_t = (\tau_t, \zeta_t, \bm{\beta}_t, \bm{\gamma}_t)$ with $\bm{\beta}_t = (\beta_{1t}, \dots, \beta_{pt})$,  $\bm{\gamma}_t = (\gamma_{1t}, \dots, \gamma_{pt})$. Since the response specific parameters $\tau_t$ and $\zeta_t$ for each unit $t$ are scalars and relatively cheap to update compared to length-p vectors such as $\bm{\beta}_t$ and $\bm{\gamma}_t$, we do not explicitly implement the partial-update of $\tau_t$ and $\zeta_t$. Partial updates with the adaptive-focus strategy are only implemented for $\bm{\beta}_t$ and $\bm{\gamma}_t$.

 \textcite{ruffieux_global-local_2020} also introduce an annealing scheme to the CAVI algorithm, which encourages broader exploration in the early stage of the algorithm and helps reach a higher local maximum of the ELBO. During this stage, a positive regularization parameter $T$ (the \emph{temperature}) multiplies the entropy term in the ELBO as in (\ref{formula:elbo}), encouraging exploration of the parameter space and reducing the risk of entrapment in poor local modes. We emphasize that the annealing phase modifies the variational objective only during the first few iterations; once the temperature reaches $T=1$, the algorithm reverts to standard CAVI (or AF-CAVI) on the unmodified ELBO. Convergence is therefore monitored on the target ELBO after the annealing phase has ended. 


Here we follow the default prior specification, initializations and annealing scheme implemented by \textcite{ruffieux_global-local_2020} in the R package \texttt{atlasqtl}. The convergence tolerance, i.e., the difference of ELBO in two consecutive iterations is set to 0.01. The annealing temperature follows the default geometric schedule with initial temperature 2 with grid 10, i.e., over the first 10 iterations the temperature $T$ is gradually reduced from 2 to 1 according to a geometric cooling schedule. To specify the prior hyperparameters $n_0$ and $t_0$, \textcite{ruffieux_global-local_2020} transfers them to more interpretable hyperparameters representing the prior mean and variance for the number of SNPs associated with each protein, namely $m_0, v_0$ respectively. We follow their default specification of $m_0 = 1$ and $v_0 = 4$, respectively, when running on simulated data (Section 3.4), and set them to be proportional to the number of SNPs ($p$) when running on blocks of different sizes on chromosome 1 (Section 3.5) with the following formula:
\[
\begin{aligned}
m_0 &= 0.001 \times p \\
v_0 &= (0.0015 \times p)^2,
\end{aligned}
\]
which corresponds to an average of 1 associated SNP per protein with variance 2.25 in a block of $p = 1\,000$. We note that during the annealing phase the variational objective is modified by the temperature parameter and is therefore not the target ELBO; consistently, the ELBO is not evaluated in this phase. Once $T=1$ is reached, AF-CAVI reduces to block coordinate ascent on the unmodified ELBO, in which any local factors not selected at a given iteration correspond to null updates with zero improvement. The sequence of ELBO values generated by AF-CAVI from this point onward is therefore non-decreasing, matching the monotonicity property of standard CAVI. As with standard CAVI, the resulting fixed point is a local optimum of the (non-convex) variational objective rather than a global one.


\clearpage
\section{Data simulation settings} \label{append:simulation}

Here we explain the details of generating additive responses $\bm{y} = (\bm{y}_1, ..., \bm{y}_q)$ from real SNPs $\bm{X} = (\bm{X}_1, ..., \bm{X}_p)$ according to
\begin{equation} \label{formula:sim}
    \bm{y}_t =  \sum_{s=1}^p \beta_{st} \bm{X}_s+ \bm{\varepsilon}_t , \ t = 1, ..., q .
\end{equation}
Our simulation follows three steps:
\begin{enumerate}
    \item Define the dependence structure $\gamma_{st} = \mathbbm{1} \{\bm{X}_s \text{ is associated with } \bm{y}_t \}$ for each pair of $\bm{X}_s$, $\bm{y}_t$;
    \item Simulate the error terms $\bm{\varepsilon}_t$ with a specified correlation structure in the responses;
    \item Simulate the effect sizes $\beta_{st}$.
\end{enumerate}
The rest of this section explains the details of each step utilizing functions in the echoseq package \cite{ruffieux_echoseq_nodate} and parameters selected according to \textcite{ruffieux_fully_2020}. No missing value is inserted in the simulated responses for simplicity.

\paragraph{Step 1: Defining the dependence structure}
Randomly select $a_p$ percentage of all SNPs to be ``active'', i.e., associated with at least one response, while the other SNPs are set as ``inactive'' and not associated with any response. For each active predictor $\bm{X}_s$, we assign a propensity parameter: 
\begin{equation} \nonumber
    \pi_s \sim \text{Beta} (1, 5),
\end{equation} 
which describes the probability of each active $\bm{X}_s$ to be associated with any active $\bm{y}_t$. The propensity parameter follows a right-skewed Beta distribution that favors small values, so that most active SNPs are assigned only one associated protein, while a small fraction of active SNPs are hotspots. In case a response is not associated with any predictor, we randomly assign one predictor from the active set to be associated with this response. The same check is applied to all active predictors. 

\paragraph{Step 2: Simulation of equicorrelated error terms}
The correlation structure of responses is defined in a grouped fashion. We first group responses into $B = \lceil q/s_q \rceil$ blocks of size $s_q = 50$. Responses within each block are equicorrelated with correlation $\eta_b$ while those from different groups are independent. The correlation  $\eta_b$ of each block is uniformly distributed between 0 and 0.5, i.e., \begin{equation}
    \eta_b \overset{\mathrm{iid}}{\sim} \text{Unif }(0, 0.5), \quad b = 1, ...,  B. \nonumber
\end{equation} 
The $s_q \times s_q$ correlation matrix $\bm{\Sigma}^{(b)}$ for responses in block $b$ is then defined as:
\begin{equation}
    \bm{\Sigma}_{ij}^{(b)} = 
    \left\{
    \begin{array}{ll}
        1, & \text{if } i = j, \\ 
        \eta_b, & \text{if } i \neq j ,
    \end{array}
    \right. 
    \nonumber 
\end{equation}
In order to simulate the error terms $\bm{\varepsilon}^{(b)} \sim N(0, \bm{\Sigma}^{(b)})$, we perform Cholesky decomposition on $\bm{\Sigma}^{(b)}$ such that $\bm{\Sigma}^{(b)} = \bm{L}^{(b)} {\bm{L}^{(b)}}^T$. Let $\bm{Z}^{(b)}$ be a $n \times s_q$ matrix of $s_q$ uncorrelated random variables normally distributed with mean 0 and variance 1. 
Then
\begin{equation}
    \bm{\varepsilon}^{(b)} = \bm{Z}^{(b)} \bm{L}^{(b)}. \nonumber
\end{equation}
follows a multivariate normal distribution with mean 0 and variance matrix $\bm{\Sigma}^{(b)}$. 

\paragraph{Step 3: Simulation of effect sizes} Effect sizes are simulated from pre-defined heritability levels. By fitting a linear regression model considering only the additive effects as described in~\ref{formula:sim}, the narrow-sense heritability (i.e., heritability solely due to the additive genetic effects, or the variance explained by $\bm{X}$) for trait $\bm{y}_t$ is 
\begin{equation}\nonumber
    h^2_{t} := \frac{\text{Var}(\bm{X} \bm{\beta}_{t} \mid \bm{\beta}_{t})}{\text{Var}(\bm{y}_t)}.
\end{equation}
Assuming uncorrelated predictors for simplification, $h^2_t$ decomposes as a sum of $h^2_{st}$, the variance of $\bm{y}_t$ explained by each predictor $\bm{X}_s$:
\begin{equation}\nonumber
     h^2_{t} = \sum_{s=1}^{p} h^2_{st}, 
\end{equation}
where 
\begin{equation} \nonumber
\begin{aligned}
    h^2_{st} &= \frac{Var(\bm{X}_s \beta_{st} \mid \beta_{st})}{\text{Var}(\bm{y}_t)} =  \frac{\beta_{st}^2 \text{Var}(\bm{X}_s)}{\text{Var}(\bm{y}_t)}. \nonumber
\end{aligned}
\end{equation}
We estimate $\text{Var}(\bm{X}_s)$ empirically as 
\begin{equation}
    \widehat{\text{Var}}(\bm{X}_s) = 2 f_s (1 - f_s),  \nonumber
\end{equation} 
where $f_s$ denotes the empirical minor allele frequency (MAF) of $\bm{X}_s$. Further, we use $\widehat{\text{Var}}(\bm{y}_t) =  \widehat{\text{Var}}(\bm{\varepsilon}_t) / (1 - h^2_t)$, where $\widehat{\text{Var}}(\bm{\varepsilon}_t)$ is estimated empirically from our simulated values in Step 2, assuming genetic factors uncorrelated from environmental factors. Thus, given $h^2_{st}$, $\beta_{st}$ can be derived as:
\begin{equation} \nonumber
    \beta_{st} = \sqrt{\frac{h^2_{st}}{1 - h^2_t} \frac{\widehat{\text{Var}}(\bm{\varepsilon}_t)}{2f_s(1-f_s)}}.
\end{equation}
Therefore, one can simulate effect size $\beta_{st}$ by specifying the proportion of variance in $\bm{y}_t$ explained by $\bm{X}_s$. Here we first select an \emph{average heritability level} $h^2_m$ for $h^2_t$, i.e.,
\[
h^2_t \sim \text{Beta}\left(1, \frac{1 - h^2_m}{h^2_m}\right).
\]
This ensures that $h^2_m$ is the expected total heritability of each response. Then for each $t$, we simulate $h^2_{st} \sim  \text{Beta}(2,5) $, and rescale $h^2_{1t}, \ldots h^2_{pt}$ such that their sum is $h^2_t$.

\clearpage
\section{Sensitivity analysis to the number of initial full updates before the adaptive-focus stage}

To evaluate the sensitivity to the number of full updates before entering the adaptive-focus stage of the algorithm (denoted by $\text{I}_{init}$), we conducted a simulation study based on 50 independently generated datasets with parameters $p = 1000, q = 3000, h_m^2 = 0.15, a_p = 0.01, a_q = 0.1$. We considered $\text{I}_{init} = \{0, 15, 30, 50, 100\}$, and compared computational efficiency and statistical performance across settings. All experiments were carried out using the AFIO-CAVI implementation with $\alpha = 0.95$. For each dataset, relative differences in runtime and total number of iterations were computed with respect to the corresponding Vanilla CAVI run on the same dataset.

\begin{figure}[h]
    \centering
    \includegraphics[width=0.8\linewidth]{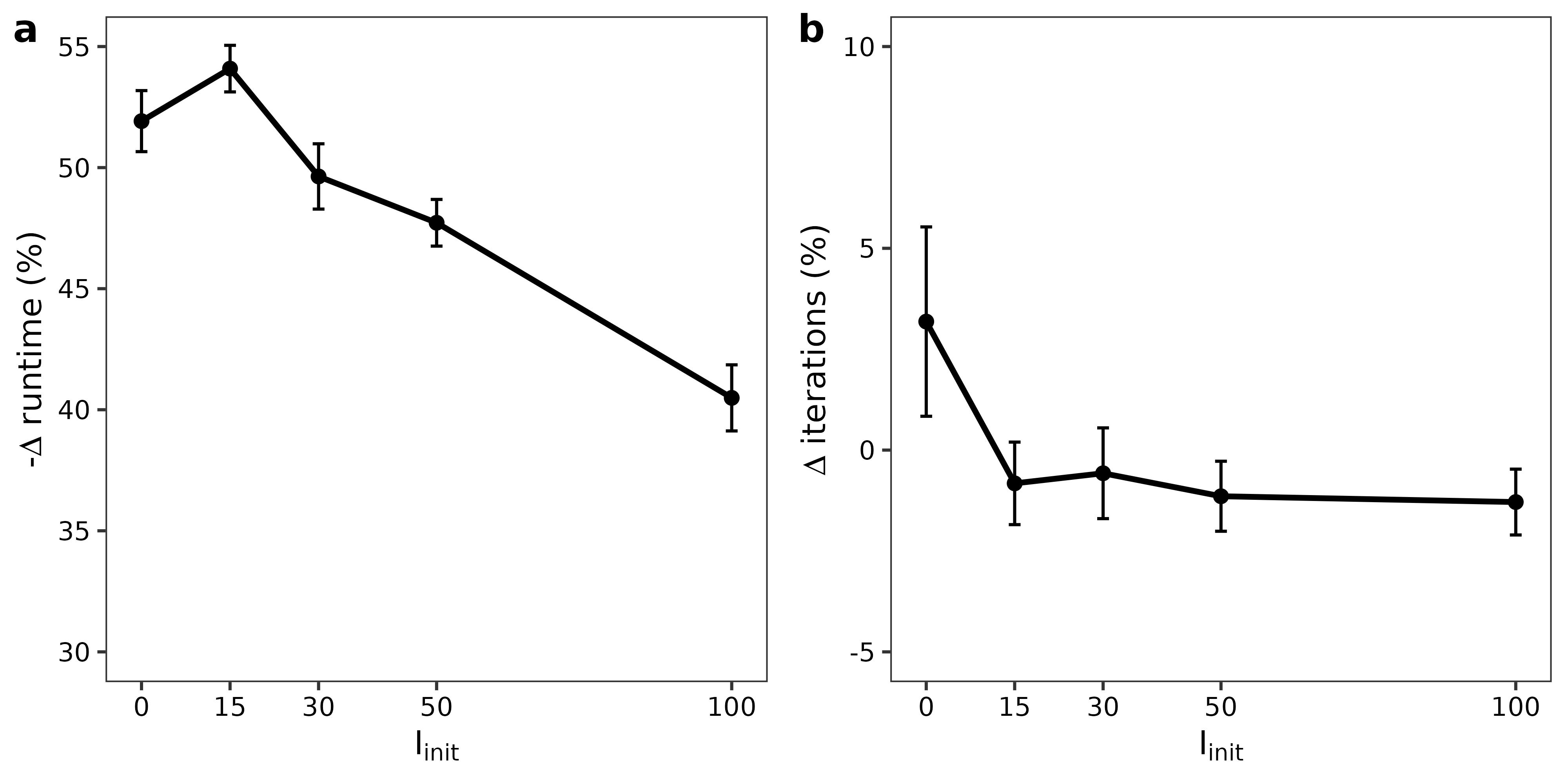}
    \caption{\textbf{Sensitivity of AF-CAVI to the number of initial full updates} (a) $-\Delta\text{runtime}$; (b) $\Delta\text{iterations}$; both computed with respect to vanilla CAVI. Results are averaged over 50 simulated datasets for $\text{I}_{init} = \{0, 15, 30, 50, 100\}$, with error bars indicating standard errors. Positive values of $-\Delta \text{runtime}$ correspond to shorter runtime relative to vanilla CAVI, while positive values of $\Delta \text{iterations}$show larger number of total iterations.}
    \label{fig:init_efficiency}
\end{figure}

Figure~\ref{fig:init_efficiency} shows that compared to entering the adaptive-focus stage directly  ($\text{I}_{init}=0$) , incorporating an initial phase of full updates ($\text{I}_{init}>0$) mitigates the increase in the number of iterations associated with adaptive focusing. However, this reduction is not very sensitive to further increases in $\text{I}_{init}$ beyond a moderate level, indicating diminishing returns from additional full updates. The statistical performance also shows little sensitivity to different $\text{I}_{init}$ values (See Table~\ref{tab:init_stats}). Thus, in this study, we use $\text{I}_{init} = 15$.

\begin{table}[t!]
\centering
\small
\begin{tabular}{lccc}
\toprule
Method & Precision & Recall & FPR ($\times 10^{-5}$) \\
\midrule
Vanilla CAVI
& $0.71 \pm 0.02$
& $0.86 \pm 0.01$
& $6.64 \pm 0.55$ \\

\midrule
AFIO-CAVI ($I_{\text{init}} = 0$)
& $0.72 \pm 0.02$
& $0.87 \pm 0.01$
& $6.55 \pm 0.62$ \\

AFIO-CAVI ($I_{\text{init}} = 15$)
& $0.71 \pm 0.02$
& $0.86 \pm 0.01$
& $6.67 \pm 0.55$ \\

AFIO-CAVI ($I_{\text{init}} = 30$)
& $0.71 \pm 0.02$
& $0.86 \pm 0.01$
& $6.65 \pm 0.55$ \\

AFIO-CAVI ($I_{\text{init}} = 50$)
& $0.71 \pm 0.02$
& $0.86 \pm 0.01$
& $6.66 \pm 0.55$ \\

AFIO-CAVI ($I_{\text{init}} = 100$)
& $0.71 \pm 0.02$
& $0.86 \pm 0.01$
& $6.66 \pm 0.55$ \\
\bottomrule
\end{tabular}
\caption{\textbf{Mean and standard error of precision, recall, and false positive rate (FPR) .}
The first row corresponds to Vanilla CAVI, and the remaining rows correspond to AFIO-CAVI with different values of $I_{\text{init}}$. FPR values are reported in units of $10^{-5}$. Average metrics are reported with ± standard error computed across 50 replicates}
\label{tab:init_stats}
\end{table}

\clearpage
\section{Sensitivity analysis to the decreasing rate of the mixing parameter in the adaptive-focus stage}

\begin{figure}[ht]
  \centering
    \includegraphics[width=0.5
    \linewidth]{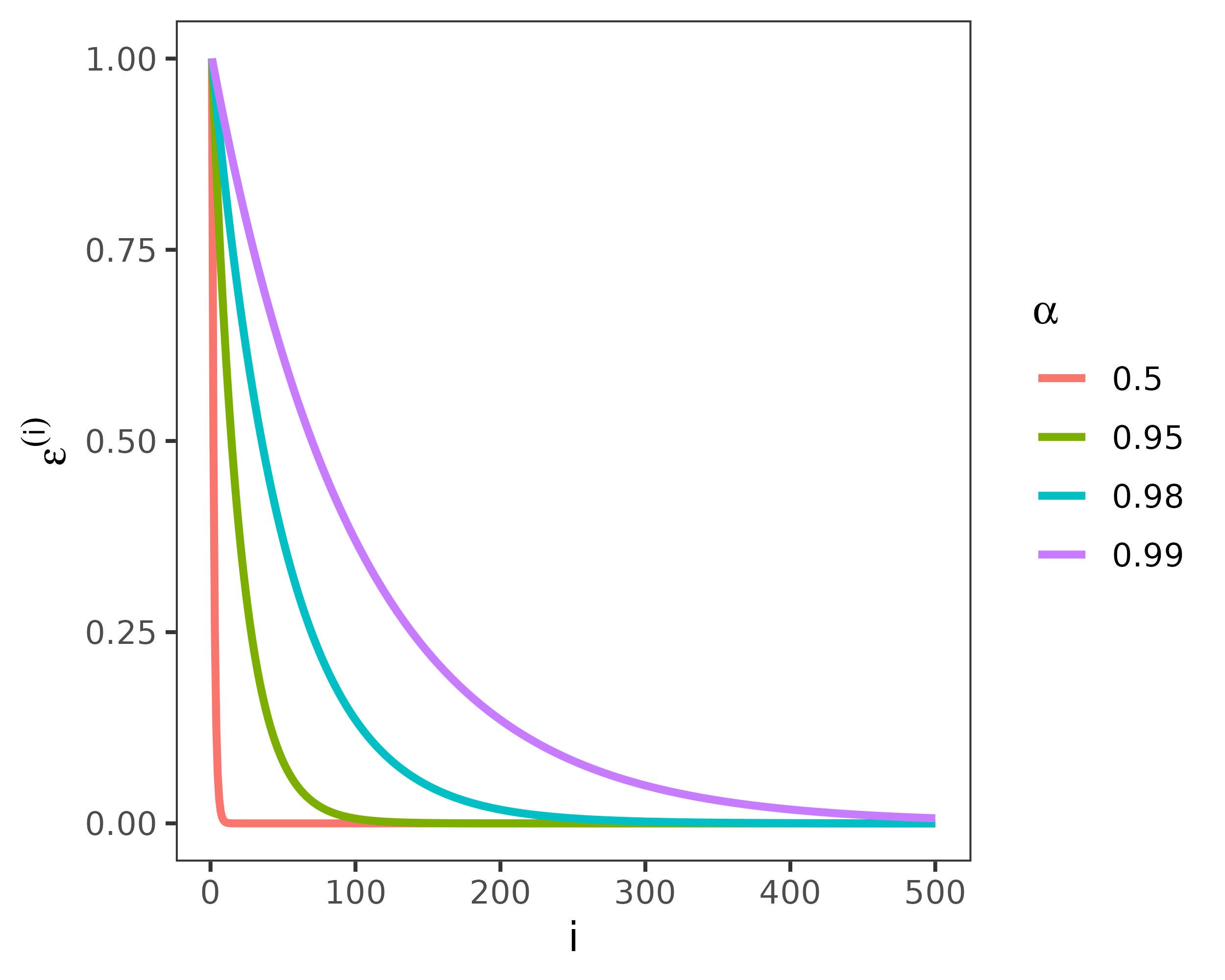}
    \caption{\textbf{Geometric decreasing scheme of the mixing parameter with different rate $\alpha$.} $x$-axis iteration index $i$. $y$-axis: the mixing parameter at iteration $i$, $\varepsilon^{(i)}$. Different decreasing schemes are distinguished by different colors over a scale of iteration $i = 0$ till $i=500$}
    \label{fig:geom_alpha_param}
\end{figure}

\begin{figure}[!htbp]
    \centering
    \includegraphics[width=0.95\linewidth]{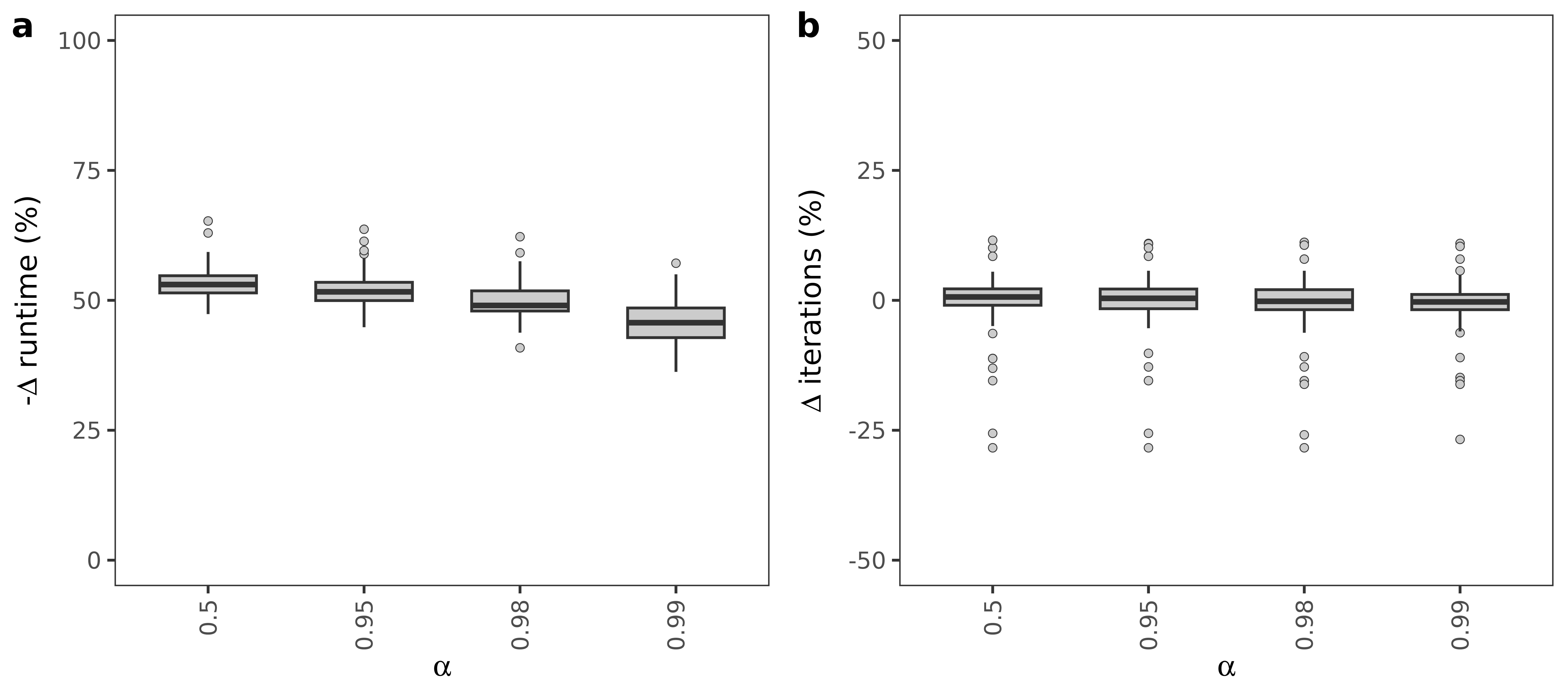}
    \caption{\textbf{Sensitivity of AF-CAVI to the decreasing rate of the mixing parameter} (a) $-\Delta\text{runtime}$; (b) $\Delta\text{iterations}$; both computed with respect to vanilla CAVI. Results are averaged over 50 simulated datasets for $\alpha = \{0.5, 0.95, 0.98, 0.99\}$. Positive values of $-\Delta \text{runtime}$ correspond to shorter runtime relative to vanilla CAVI, while positive values of $\Delta \text{iterations}$ show larger number of total iterations.}
    \label{fig:geom_alpha}
\end{figure}

In this section we demonstrate the sensitivity analysis of the geometric decreasing rate $\alpha$ of the mixing parameter $\varepsilon^{(i)}$ (with $i$ denoting the index of iteration) described in Section 3.3,
$\varepsilon^{(i)} = \alpha^{i-1} .$
The mixing parameter decreases from 1 to 0, and controls the degree to which the selection process is driven by the activity score at iteration $i$. Larger values of $\varepsilon^{(i)}$ will let the algorithm select a larger number of local factors in early iterations, while smaller values restrict updates to fewer factors that are more likely to be active. Here we simulate 50 independently generated datasets with parameters $p = 1000, q = 3000, h_m^2 = 0.15, a_p = 0.01, a_q = 0.1$, and run AFIO-CAVI with $\alpha = \{0.5, 0.95, 0.98, 0.99\}$ (See Figure~\ref{fig:geom_alpha_param} for an illustration of the geometric decreasing scheme with different $\alpha$). Relative differences in runtime and number of iterations are computed with respect to Vanilla CAVI (Figure~\ref{fig:geom_alpha}). 

Overall, the algorithm shows little sensitivity to $\alpha$. Relatively moderate values such as 0.95, 0.98, 0.99 show very little difference in terms of runtime and number of iterations. The difference in statistical performance among different choices of $\alpha$ is also marginal (Table~\ref{tab:alpha_stats}). Therefore a default value of  $\alpha = 0.95$ is applied throughout this study.

\begin{table}[ht]
\centering
\small
\begin{tabular}{lccc}
\toprule
Method & Precision & Recall & FPR ($\times 10^{-5}$) \\
\midrule
Vanilla CAVI
& $0.71 \pm 0.02$
& $0.86 \pm 0.01$
& $6.64 \pm 0.55$ \\

\midrule
AFIO-CAVI ($\alpha = 0.50$)
& $0.71 \pm 0.02$
& $0.86 \pm 0.01$
& $6.68 \pm 0.55$ \\

AFIO-CAVI ($\alpha = 0.95$)
& $0.71 \pm 0.02$
& $0.86 \pm 0.01$
& $6.67 \pm 0.55$ \\

AFIO-CAVI ($\alpha = 0.98$)
& $0.71 \pm 0.02$
& $0.86 \pm 0.01$
& $6.68 \pm 0.55$ \\

AFIO-CAVI ($\alpha = 0.99$)
& $0.71 \pm 0.02$
& $0.86 \pm 0.01$
& $6.68 \pm 0.55$ \\
\bottomrule
\end{tabular}
\caption{\textbf{Mean and standard error of precision, recall, and false positive rate (FPR) for different values of the mixing parameter $\alpha$.}
The first row corresponds to Vanilla CAVI. The remaining rows correspond to AFIO-CAVI with different values of $\alpha$. FPR values are reported in units of $10^{-5}$.  Average metrics are reported with ± standard error computed across 50 replicates. }
\label{tab:alpha_stats}
\end{table}

\clearpage
\section{Simulation results across all scenarios} \label{append:full res}
This following table shows the simulation result across all simulation scenarios with different heritability levels ($h_m^2 = 0.05, 0.15, 0.3$) and sparsity levels ($a_q = 0.005, 0.2, 0.5$)  , as supplementary to the result shown for the case of $h_m = 0.15$. This table shows that the relative performance of different methods is not sensitive to different heritability or sparsity levels.

\begin{table}[!h]
    \centering
    \scriptsize
\begin{tabular}[t]{rrlllllll}
\toprule
$h_m^2$ & $a_q$ & Method & $\Delta$ Iterations \% & $-\Delta$ Runtime (total) \% & $-\Delta$ Runtime (local) \% & FPR ($\times 10^{-4}$) & Precision & Recall\\
\midrule
\multirow{15}{*}{\raggedleft\arraybackslash 0.05} & \multirow{5}{*}{\raggedleft\arraybackslash 0.005} & Vanilla & -- & -- & -- & $0.04 \pm 0.0033$ & $0.67 \pm 0.02$ & $0.88 \pm 0.02$ \\
 &  & RF & $0.47 \pm 0.30$ & $34.24 \pm 4.24$ & $45.63 \pm 4.74$ & $0.04 \pm 0.0034$ & $0.67 \pm 0.02$ & $0.88 \pm 0.02$ \\
 &  & AFE & $0.07 \pm 0.07$ & $47.64 \pm 3.20$ & $63.00 \pm 3.03$ & $0.04 \pm 0.0033$ & $0.67 \pm 0.02$ & $0.88 \pm 0.02$ \\
 &  & AFI & $1.49 \pm 0.05$ & $57.27 \pm 3.53$ & $79.49 \pm 2.35$ & $0.04 \pm 0.0034$ & $0.67 \pm 0.02$ & $0.88 \pm 0.02$ \\
 &  & AFIO & $1.49 \pm 0.05$ & $69.72 \pm 1.63$ & $85.32 \pm 0.99$ & $0.04 \pm 0.0034$ & $0.67 \pm 0.02$ & $0.88 \pm 0.02$ \\
  \cmidrule{2-9}
 & \multirow{5}{*}{\raggedleft\arraybackslash 0.2} & Vanilla & -- & -- & -- & $1.41 \pm 0.11$ & $0.68 \pm 0.02$ & $0.79 \pm 0.01$ \\
 &  & RF & $11.04 \pm 1.72$ & $24.37 \pm 4.81$ & $39.86 \pm 5.25$ & $1.42 \pm 0.11$ & $0.68 \pm 0.02$ & $0.79 \pm 0.01$ \\
 &  & AFE & $0.08 \pm 0.31$ & $25.57 \pm 3.92$ & $39.62 \pm 4.27$ & $1.41 \pm 0.11$ & $0.68 \pm 0.02$ & $0.79 \pm 0.01$ \\
 &  & AFI & $0.38 \pm 0.21$ & $30.64 \pm 3.50$ & $53.78 \pm 3.29$ & $1.41 \pm 0.11$ & $0.68 \pm 0.02$ & $0.79 \pm 0.01$ \\
 &  & AFIO & $0.38 \pm 0.21$ & $48.23 \pm 3.04$ & $64.59 \pm 2.77$ & $1.41 \pm 0.11$ & $0.68 \pm 0.02$ & $0.79 \pm 0.01$ \\
  \cmidrule{2-9}
 & \multirow{5}{*}{\raggedleft\arraybackslash 0.5} & Vanilla & -- & -- & -- & $3.47 \pm 0.30$ & $0.70 \pm 0.02$ & $0.81 \pm 0.01$ \\
 &  & RF & $26.62 \pm 2.86$ & $18.91 \pm 2.18$ & $37.35 \pm 2.25$ & $3.50 \pm 0.30$ & $0.70 \pm 0.02$ & $0.81 \pm 0.01$ \\
 &  & AFE & $0.12 \pm 0.52$ & $9.90 \pm 4.92$ & $17.75 \pm 6.31$ & $3.48 \pm 0.30$ & $0.70 \pm 0.02$ & $0.81 \pm 0.01$ \\
 &  & AFI & $1.12 \pm 0.87$ & $23.66 \pm 1.84$ & $37.75 \pm 2.29$ & $3.47 \pm 0.30$ & $0.70 \pm 0.02$ & $0.81 \pm 0.01$ \\
 &  & AFIO & $1.12 \pm 0.87$ & $33.90 \pm 1.37$ & $42.53 \pm 1.69$ & $3.47 \pm 0.30$ & $0.70 \pm 0.02$ & $0.81 \pm 0.01$ \\
\cmidrule{1-9}
\multirow{15}{*}{\raggedleft\arraybackslash 0.15} & \multirow{5}{*}{\raggedleft\arraybackslash 0.005} & Vanilla & -- & -- & -- & $0.04 \pm 0.0034$ & $0.73 \pm 0.02$ & $0.92 \pm 0.01$ \\
 &  & RF & $0.20 \pm 0.51$ & $32.55 \pm 0.84$ & $50.28 \pm 0.83$ & $0.04 \pm 0.0034$ & $0.70 \pm 0.02$ & $0.92 \pm 0.01$ \\
 &  & AFE & $-0.11 \pm 0.16$ & $33.53 \pm 0.78$ & $53.44 \pm 0.79$ & $0.04 \pm 0.0034$ & $0.73 \pm 0.02$ & $0.92 \pm 0.01$ \\
 &  & AFI & $1.22 \pm 0.17$ & $48.71 \pm 0.52$ & $78.10 \pm 0.29$ & $0.04 \pm 0.0034$ & $0.73 \pm 0.02$ & $0.92 \pm 0.01$ \\
 &  & AFIO & $1.22 \pm 0.17$ & $56.26 \pm 0.48$ & $78.47 \pm 0.32$ & $0.04 \pm 0.0034$ & $0.73 \pm 0.02$ & $0.92 \pm 0.01$ \\
  \cmidrule{2-9}
 & \multirow{5}{*}{\raggedleft\arraybackslash 0.2} & Vanilla & -- & -- & -- & $1.31 \pm 0.12$ & $0.72 \pm 0.02$ & $0.86 \pm 0.01$ \\
 &  & RF & $37.09 \pm 2.63$ & $14.07 \pm 1.91$ & $32.79 \pm 1.87$ & $1.35 \pm 0.12$ & $0.71 \pm 0.02$ & $0.87 \pm 0.01$ \\
 &  & AFE & $0.15 \pm 0.77$ & $25.54 \pm 1.35$ & $39.20 \pm 1.63$ & $1.31 \pm 0.12$ & $0.72 \pm 0.02$ & $0.86 \pm 0.01$ \\
 &  & AFI & $-0.42 \pm 0.93$ & $42.40 \pm 0.83$ & $65.65 \pm 0.71$ & $1.31 \pm 0.12$ & $0.72 \pm 0.02$ & $0.86 \pm 0.01$ \\
 &  & AFIO & $-0.42 \pm 0.93$ & $50.92 \pm 0.62$ & $66.82 \pm 0.60$ & $1.31 \pm 0.12$ & $0.72 \pm 0.02$ & $0.86 \pm 0.01$ \\
  \cmidrule{2-9}
 & \multirow{5}{*}{\raggedleft\arraybackslash 0.5} & Vanilla & -- & -- & -- & $3.23 \pm 0.31$ & $0.74 \pm 0.02$ & $0.88 \pm 0.01$ \\
 &  & RF & $46.19 \pm 3.12$ & $4.43 \pm 3.11$ & $24.54 \pm 2.96$ & $3.27 \pm 0.31$ & $0.73 \pm 0.02$ & $0.88 \pm 0.01$ \\
 &  & AFE & $1.32 \pm 1.19$ & $17.51 \pm 1.53$ & $27.85 \pm 1.69$ & $3.22 \pm 0.31$ & $0.74 \pm 0.02$ & $0.88 \pm 0.01$ \\
 &  & AFI & $1.65 \pm 1.78$ & $31.37 \pm 1.46$ & $47.73 \pm 1.37$ & $3.22 \pm 0.31$ & $0.74 \pm 0.02$ & $0.88 \pm 0.01$ \\
 &  & AFIO & $1.65 \pm 1.78$ & $36.57 \pm 1.32$ & $45.11 \pm 1.37$ & $3.22 \pm 0.31$ & $0.74 \pm 0.02$ & $0.88 \pm 0.01$ \\
\cmidrule{1-9}
\multirow{15}{*}{\raggedleft\arraybackslash 0.3} & \multirow{5}{*}{\raggedleft\arraybackslash 0.005} & Vanilla & -- & -- & -- & $0.04 \pm 0.0037$ & $0.69 \pm 0.02$ & $0.94 \pm 0.01$ \\
 &  & RF & $-2.44 \pm 1.30$ & $38.92 \pm 1.88$ & $55.17 \pm 1.88$ & $0.06 \pm 0.0047$ & $0.62 \pm 0.02$ & $0.94 \pm 0.01$ \\
 &  & AFE & $0.21 \pm 0.42$ & $37.06 \pm 1.87$ & $55.17 \pm 1.96$ & $0.04 \pm 0.0038$ & $0.69 \pm 0.02$ & $0.94 \pm 0.01$ \\
 &  & AFI & $2.47 \pm 0.42$ & $55.04 \pm 1.09$ & $81.85 \pm 0.60$ & $0.04 \pm 0.0036$ & $0.69 \pm 0.02$ & $0.94 \pm 0.01$ \\
 &  & AFIO & $2.47 \pm 0.42$ & $61.41 \pm 0.95$ & $81.82 \pm 0.61$ & $0.04 \pm 0.0036$ & $0.69 \pm 0.02$ & $0.94 \pm 0.01$ \\
  \cmidrule{2-9}
 & \multirow{5}{*}{\raggedleft\arraybackslash 0.2} & Vanilla & -- & -- & -- & $1.47 \pm 0.13$ & $0.70 \pm 0.02$ & $0.91 \pm 0.01$ \\
 &  & RF & $38.30 \pm 3.25$ & $-7.70 \pm 7.27$ & $7.67 \pm 8.30$ & $1.54 \pm 0.13$ & $0.69 \pm 0.02$ & $0.91 \pm 0.01$ \\
 &  & AFE & $-0.58 \pm 0.91$ & $23.70 \pm 3.67$ & $35.78 \pm 4.36$ & $1.47 \pm 0.13$ & $0.70 \pm 0.02$ & $0.91 \pm 0.01$ \\
 &  & AFI & $-0.23 \pm 1.00$ & $40.62 \pm 2.51$ & $63.17 \pm 2.36$ & $1.47 \pm 0.13$ & $0.70 \pm 0.02$ & $0.91 \pm 0.01$ \\
 &  & AFIO & $-0.23 \pm 1.00$ & $56.24 \pm 0.64$ & $71.24 \pm 0.52$ & $1.47 \pm 0.13$ & $0.70 \pm 0.02$ & $0.91 \pm 0.01$ \\
  \cmidrule{2-9}
 & \multirow{5}{*}{\raggedleft\arraybackslash 0.5} & Vanilla & -- & -- & -- & $3.63 \pm 0.34$ & $0.72 \pm 0.02$ & $0.92 \pm 0.01$ \\
 &  & RF & $39.79 \pm 3.27$ & $8.24 \pm 3.13$ & $27.21 \pm 3.14$ & $3.72 \pm 0.34$ & $0.72 \pm 0.02$ & $0.92 \pm 0.01$ \\
 &  & AFE & $0.33 \pm 0.82$ & $18.23 \pm 1.41$ & $27.92 \pm 1.39$ & $3.61 \pm 0.34$ & $0.72 \pm 0.02$ & $0.92 \pm 0.01$ \\
 &  & AFI & $-0.33 \pm 0.93$ & $21.98 \pm 3.62$ & $35.86 \pm 4.19$ & $3.62 \pm 0.34$ & $0.72 \pm 0.02$ & $0.92 \pm 0.01$ \\
 &  & AFIO & $-0.33 \pm 0.93$ & $28.41 \pm 4.25$ & $32.56 \pm 5.57$ & $3.62 \pm 0.34$ & $0.72 \pm 0.02$ & $0.92 \pm 0.01$ \\
\bottomrule
\end{tabular}
\captionof{table}{\textbf{Performance of Vanilla, RF, and different implementations of AF-CAVI in different simulation scenarios.} Results are shown for heritability levels $h_m^2 = 0.05, 0.15, 0.3$ and sparsity levels $a_q = 0.005, 0.2, 0.5$ with 50 independently simulated datasets. For RF, AFE, AFI and AFIO, relative change w.r.t. Vanilla CAVI in iterations ($\Delta$ Iterations) and (negative) relative change in total and local runtime ($-\Delta$ Runtime (total) and $-\Delta$ Runtime (local)) are shown in percentage. For all methods, we show the absolute values of FPR, precision and recall. Values are shown in format mean $\pm$ standard error.}
\end{table}

\clearpage
\section{List of associated proteins of SNP rs2476601 reported by BHM and univariate testing}

This following table shows the list of associated proteins of SNP rs2476601 located in the \emph{PTPN22} gene reported by BHM and univariate testing. BHM reports 10 additional proteins compared to univariate testing. In the 6 proteins that both BHM and univariate testing reported as significant, the regression coefficients are consistent. 
\begin{figure}[!htbp]
\centering
\scriptsize
\begin{tabular}[t]{cccccccc}
\toprule
Protein assay & Chromosome & TSS (bp) & cis or trans & PPI (BHM) & $-log_{10}(\text{\emph{p}-value)}$ (Univariate) & BETA (BHM) & BETA (Univariate)\\
\midrule
CCL19 & 9 & 34691274 & trans & 1.0000000 & 13.60644 & 0.1250253 & 0.1258630\\
CXCL9 & 4 & 76928641 & trans & 1.0000000 & 12.70917 & 0.0652627 & 0.0654865\\
CHAD & 17 & 48542786 & trans & 0.9999999 & 12.36139 & -0.0445743 & -0.0448394\\
COL1A1 & 17 & 48278993 & trans & 0.9999994 & 11.41852 & -0.0233178 & -0.0234838\\
IL10 & 1 & 206945839 & trans & 0.9999986 & 11.04570 & 0.0733044 & 0.0737567\\
PDCD1 & 2 & 242801060 & trans & 0.9999975 & 10.79505 & 0.0382918 & 0.0385567\\
IL12B & 5 & 158757895 & trans & 0.9999650 & NA & 0.0480400 & NA\\
CD5L & 1 & 157811588 & trans & 0.9999553 & NA & -0.0366879 & NA\\
CXCL10 & 4 & 76944650 & trans & 0.9999474 & NA & 0.0565372 & NA\\
IL12A\_IL12B & 3 & 159706537 & trans & 0.9997745 & NA & 0.0643592 & NA\\
LIPF & 10 & 90424198 & trans & 0.9993131 & NA & -0.0428626 & NA\\
MMP13 & 11 & 102826463 & trans & 0.9990262 & NA & -0.0239698 & NA\\
CRTAM & 11 & 122709208 & trans & 0.9985055 & NA & 0.0383698 & NA\\
CD5 & 11 & 60869867 & trans & 0.9982527 & NA & 0.0261474 & NA\\
JCHAIN & 4 & 71532377 & trans & 0.9971074 & NA & -0.0450441 & NA\\
SPP1 & 4 & 88896819 & trans & 0.9918606 & NA & -0.0324990 & NA\\
\bottomrule
\end{tabular}
    \caption{\textbf{List of associated proteins of SNP rs2476601 reported by BHM and univariate testing.} Column Protein assay: measured plasma protein. Chromosome: the chromosome on which the gene encoding the protein is located. TSS (bp) gives the genomic position (base pairs, hg19) of the transcription start site of the corresponding gene. cis or trans: classifies the association based on genomic distance to the SNP, where \textit{cis} indicates the SNP lies within $\pm 0.5$ Mb of the protein’s TSS and \textit{trans} indicates the SNP lies outside this window (including variants on different chromosomes). PPI (BHM): posterior probability of inclusion estimated by the Bayesian hierarchical model (atlasQTL). –$\log_{10}(\text{\emph{p}-value})$ (Univariate): significance level from single-protein univariate association testing. BETA (BHM) and BETA (Univariate): estimated effect sizes from the BHM and univariate models respectively. NA indicates that the association did not meet the significance threshold in univariate testing.
}
    \label{tab:placeholder}
\end{figure}

\end{document}